\documentclass[twocolumn,epjc3]{svjour3}  
\smartqed
\usepackage[utf8,latin1]{inputenc}
\newcommand{\qm}[1]{``#1''}
\usepackage{amsmath}
\usepackage{amssymb}
\usepackage{mathabx}
\usepackage{mathrsfs}  
\usepackage{cases}
\usepackage{colortbl}
\usepackage{dcolumn}
\usepackage{xcolor}
\usepackage{accents}
\usepackage{dcolumn}
\usepackage{xcolor,colortbl}
\usepackage{multirow}
\usepackage{stackengine}
\usepackage{bm}
\usepackage{tikz-cd}
\usepackage{tikz}
\usepackage{bbm}
\usepackage{hyperref}
\hypersetup{colorlinks, linkcolor={red},citecolor={blue},urlcolor={blue}}  
\usepackage[square,numbers,sort&compress]{natbib}
\usepackage{hyperref}
\usepackage{stackengine,scalerel}

\RequirePackage{graphicx}
\journalname{Eur. Phys. J. C}
\newcommand{\Lagr}{\mathcal{L}}

\newcommand{\La}{\mathscr{L}}
\newcommand{\ee}{\end{equation}}
\newcommand{\bea}{\begin{eqnarray}}
\newcommand{\eea}{\end{eqnarray}}
\def\ba{\begin{array}}
\def\ea{\end{array}}

\newcommand\RS{{\mathrm{R_\star}}}
\newcommand{\dd}{{\rm d}}
\usepackage{enumitem}

\begin{document}

\title{Exploring departures from Schwarzschild black hole in $f(R)$ gravity}
\titlerunning{Exploring departures from Schwarzschild  black hole in $f(R)$ gravity}

\author{Vittorio~De~Falco\thanksref{e1,addr1,addr2}
\and
Francesco~Bajardi\thanksref{e2,addr1,addr2}
\and
Rocco~D'Agostino\thanksref{e3,addr1,addr2}
\and 
Micol~Benetti\thanksref{e4,addr1,addr2}
\and
Salvatore~Capozziello\thanksref{e5,addr1,addr2,addr3}}

\thankstext{e1}{e-mail: v.defalco@ssmeridionale.it}
\thankstext{e2}{e-mail: f.bajardi@ssmeridionale.it}
\thankstext{e3}{e-mail: rocco.dagostino@unina.it}
\thankstext{e4}{e-mail: micol.benetti@unina.it}
\thankstext{e5}{e-mail: capozziello@na.infn.it}

\institute{Scuola Superiore Meridionale, Largo San Marcellino 10, 80138 Napoli, Italy \label{addr1}
\and
Istituto Nazionale di Fisica Nucleare, Sezione di Napoli, Complesso Universitario di Monte S. Angelo, Via Cintia Edificio 6, 80126 Napoli, Italy \label{addr2}
\and
Dipartimento di Fisica "E. Pancini", Universit\'a di Napoli "Federico II", Complesso Universitario di Monte S. Angelo, Via Cinthia Edificio 6, I-80126 Napoli, Italy \label{addr3}}

\date{Received: \today / Accepted: }
\maketitle

\begin{abstract}
Different astrophysical methods can be combined to detect possible deviations from General Relativity. In this work, we consider a class of $f(R)$ gravity models selected by the existence of Noether symmetries. In this framework, it is possible to determine a set of static and spherically symmetric black hole solutions, encompassing small departures from the Schwarzschild geometry. In particular, when gravity is the only dominating interaction, we exploit the ray-tracing technique to reconstruct the image of a black hole, the epicyclic frequencies, and the black hole shadow profile. Moreover, when matter dynamics is also affected by an electromagnetic radiation force, we take into account the general relativistic Poynting-Robertson effect. In light of the obtained results, the proposed strategy results to be robust and efficient: on the one hand, it allows to investigate gravity from strong to weak field regimes; on the other hand, it is capable of detecting small departures from General Relativity, depending on the current observational sensitivity. 
\end{abstract}


\maketitle

\section{Introduction}
\label{sec:intro}
The recent breakthrough discoveries represented by the direct detection of gravitational waves \cite{Sathyaprakash2009,Abbott2021}, as well as the capacity to map the matter motion in the close vicinity of supermassive black holes (BHs) \cite{EHC20191,EHC20194,Akiyama2022image} have provided a large amount of complementary and sensitive observational data to inquire more deeply self-gravitating systems and gravitational interaction.  General Relativity (GR) successfully passed several observational tests (see \emph{e.g.}, Refs \cite{Ciufolini2004,Abuter2020,Kramer2021}), still confirming as the most reliable and probed theory of gravity so far available. Nevertheless, from both theoretical and observational points of view, there are critical open questions that GR is not able to address, such as the dark energy problem \cite{Carroll2001,Padmanabhan2003,Peebles2003,  DAgostino:2019wko,DAgostino:2021vvv,Perivolaropoulos:2021jda, Bamba:2012cp,DAgostino:2022fcx}, the dark matter issue \cite{Arun2017,Salucci2020}, the unification of gravity with the other fundamental interactions \cite{Ross1984} towards a full quantum description of gravity \cite{Zwiebach:2004tj}. To alleviate such shortcomings, alternatives or extensions of Einstein's theory have been proposed with the aim to relax some main assumptions of GR  or to extend it in some way
\cite{Capozziello2011,Capozziello2019,Capozziello:2022wgl,Capozziello:2022rac,DAgostino:2022tdk,Capozziello:2021goa,Capozziello:2022zzh,Bajardi:2022tzn}. However, none of the extensions or alternative models so far proposed has been proven to address all the aforementioned issues at once. It is widely believed that the strong gravitational field in the close vicinity of a BH, which has been not yet thoroughly investigated, may contain the key to uncover new physics and definitely confirm GR, or disregard it in favour of some new proposal \cite{Psaltis2020,Kramer2021,Akiyama2022image}. 
Therefore, it becomes of utmost importance to improve our insight for comparing theoretical results with observational data. Along this line of thought, some research programs resort to general BH parametrizations, where a theory-agnostic procedure is pursued \cite{Johannsen2010,Konoplya2016,Medeiros2020}. 

This approach takes into account generic metrics capable of reproducing BH geometries belonging to different gravity theories and being functions of a restricted set of parameters. This method aims at detecting departures from GR metrics, which are usually constrained through the fit of the observational data  \cite{Younsi2016,Delaurentis2018,Volkel2019}. It consists of two important steps: $(1)$ determining, through some astrophysical phenomena, whether there are significant deviations from GR; $(2)$ if the first point is accomplished, a strategy has to be developed in order to reconstruct, from the observational data, the most suitable BH solutions, which, in turn, may provide indirect tests of gravity.

In this paper, we combine different astrophysical methods to accurately determine deviations from the Schwarzschild metric. For this purpose, we consider $f(R)$ gravity as a generalization of the Hilbert-Einstein action including a generic function of the Ricci scalar, $R$ \cite{Capozziello:2002rd, Capozziello:2007ec, Nojiri2010, Sotiriou2008,DeFelice2010,Capozziello:2017ddd, Nojiri:2017ncd, DAgostino:2019hvh}. 
In particular, we exploit the \emph{Noether symmetry approach} \cite{Capozziello:1994du, Capozziello:1996bi, Dialektopoulos:2018qoe, Urban:2020lfk, Bajardi:2021tul, Bajardi:2020xfj, Bajardi:2022ypn} to select $f(R)$ power-law models such as $f(R)\sim R^k$,  with $k\in\mathbb{R}$, so that GR is recovered as soon as $k \rightarrow 1$. The same model has been considered in Ref.~\cite{Bajardi:2022ocw}, within the cosmological context, whereas here we extend the analysis to astrophysical scales. Therefore, we search for static and spherically symmetric BH solutions to be probed by astrophysical methods. 
 
As a first approximation, non-gravitational effects around  compact objects can be considered quite small with respect to the gravitational field \cite{Dovciak2003,Dauser2010}. In this scenario, our strategy relies on the ray-tracing technique to reconstruct the BH image in the observer plane \cite{Luminet1979,EHC20191,Akiyama2022image}, the epicyclic frequencies \cite{Abramowicz2005,Torok2005,Defalco2021EF}, and the BH shadow profile \cite{Falcke2000,Perlick2022}.

On the other hand, when high-energy electromagnetic processes occur around a BH (\emph{e.g.} the emission from a hot corona \cite{Fabian2015}), we exploit the relativistic Poynting-Robertson (PR) effect \cite{Poynting1903,Robertson1937,Bini2009,Bini2011}. In this context, three forces come into the game: the gravitational pull directed inwards the BH, the radiation pressure exerted outwards, and the PR radiation drag force \cite{Poynting1903,Robertson1937}. 
When radiation is absorbed by matter, treated as a test particle, the latter re-emits isotropically in its own rest frame. This produces a back-reaction force, which efficiently removes energy and angular momentum \cite{Poynting1903,Robertson1937,DeFalco2019,DeFalco2019VE}. The relativistic PR mechanism has been usually invoked to solve some puzzling issues on the dynamics of accretion disks when invested by type-I X-ray bursts \cite{Worpel2015,Keek2018,Fragile2020}.   
In addition, such an effect has been extensively studied in GR both in two and three dimensions \cite{Bini2009,Bini2011,Defalco2018,DeFalco20183D,Wielgus2019,Bakala2019,DeFalco2020,DeFalco2021HT}, featuring also chaotic behaviours under particular parameter configurations \cite{Defalco2021chaos,DeFalco2021TS}. Depending on the intensity of the electromagnetic field, if the radiation pressure dominates over the attracting forces, the test particles escape to infinity; otherwise, the PR drag force drives the test particle towards a critical hypersurface, which is located outside the BH event horizon \cite{Defalco2019ST}. Such a hypersurface is a region where all forces balance and it is characterized by stable motions of test particles \cite{DeFalco2019}. This particular feature  may represent a fundamental tool to inquiry the properties of the underlying  geometry \cite{Defalco2020wh,Defalco2021wh}.

The present paper is organized as follows. In Sect.~\ref{sec:f(R)_selction}, we introduce $f(R)$ gravity, from which we select a class of models via the Noether symmetries. Assuming a static and spherically symmetric spacetime, we single out specific BH solutions (see \cite{Bajardi:2022ypn} for details). In Sect.~\ref{sec:analysis_gravity}, we analyse deviations from Schwarzschild geometry when only gravitational interaction is present. Moreover, in Sect.~\ref{sec:PR effect}, departures from GR are studied when the electromagnetic radiation force interacts with the surrounding matter. Finally, in Sect.~\ref{sec:end}, we discuss the obtained results and outline future perspectives.

\section{Static and spherically symmetric  solutions in $ f(R)$ gravity}
\label{sec:f(R)_selction}
In this section, we select particular $f(R)$ gravity models via the Noether symmetry approach. We look for static and spherically symmetric metrics  and determine a particular BH solution via astrophysical considerations. 

\subsection{$f(R)$ gravity}
\label{sec:f(R) theory}
In $f(R)$ gravity, the Hilbert-Einstein action is generalized as follows\footnote{Throughout this work, we use units in which $c = \hbar =8 \pi G =  1$.}\cite{Nojiri2010,Capozziello:2010zz}:
\begin{equation} \label{eq:action}
    S = \int \dd^4x\, \sqrt{-g} \left[ \frac{1}{2}f(R) +\La_m\right],
\end{equation}
with $g$ being the determinant of the metric tensor $g_{\mu\nu}$ and $\La_m$ the matter Lagrangian density. By varying the action \eqref{eq:action} with respect to $g_{\mu\nu}$, one obtains the following field equations
\begin{equation} \label{eq:FE_f(R)}
f_R \, G_{\mu \nu} -\frac{1}{2} g_{\mu \nu} (f - R f_R) - \nabla_\mu\nabla_\nu f_R+ g_{\mu \nu} \Box f_R = T_{\mu \nu}\,,
\end{equation}
where $f_R\equiv \frac{\partial f}{\partial R}$. Here $G_{\mu \nu}\equiv  R_{\mu \nu} - \frac{1}{2} g_{\mu \nu} R$ is the Einstein tensor, $\nabla_\mu$ is the covariant derivative, $\Box\equiv g^{\mu\nu}\nabla_\mu\nabla_\nu$ is the d'Alembert operator, and $T_{\mu\nu}$ the energy-momentum tensor of matter fields,
\begin{equation}
T_{\mu\nu}:=-\dfrac{2}{\sqrt{-g}}\dfrac{\delta \La_m}{\delta g^{\mu\nu}}\,,
\end{equation}
satisfying the conservation law $\nabla^{\nu}T_{\mu\nu}=0$. It is worth  noticing that, for $f(R)=R$, GR is fully recovered. 

\subsection{The Noether symmetry approach}
\label{sec:NSA}
\label{sec:general}
Among all $f(R)$ extensions of GR, viable models can be selected using the so-called Noether symmetry approach \cite{Capozziello:1996bi, Capozziello:2008ch, Urban:2020lfk, Dialektopoulos:2018qoe, Acunzo:2021gqc}. This is a criterion, based on the Noether theorem, which allows to identify  models endowed with symmetries. The method consists of assuming the existence of a point transformation, which leaves the point-like Lagrangian invariant and, afterwards, in selecting the related generator $\mathcal{X}$, 
depending on the configuration variables $q^i$:
\begin{equation}
    \mathcal{X} = \eta^i \frac{\partial}{\partial q^i}\,.
\end{equation}
This can be also written in terms of the Noether vector $X$ as
\begin{equation}
    X = \mathcal{X} + \partial_\mu \eta^i  \frac{\partial}{\partial (\partial_\mu q^i)}\,,
\end{equation}
with $\eta^i$ being unknown functions of $q^i$. The first prolongation of $X$, including the first derivative transformation, is
\begin{equation}
X^{[1]} = \xi^\mu \partial_\mu + \eta^i \frac{\partial }{\partial q^i} + (\partial_\mu \eta^i - \partial_\mu q^i \partial_\nu \xi^\nu) \frac{\partial}{\partial (\partial_\mu q^i)} ,
\end{equation}
where $\xi^\mu$ is the infinitesimal generator related to the spacetime coordinate transformations. 
According to the Noether theorem, the condition
\begin{equation}
X^{[1]} \La + \partial_\mu \xi^\mu \La = \partial_\mu \mathfrak{g},
    \label{noethidentity}
\end{equation}
implies that the point transformation is a symmetry for the Lagrangian and the symmetry generator $\mathcal{X}$ automatically leads to the following integral of motion:
\begin{equation}
   j^{\mu} = -\frac{\partial \La}{\partial (\partial_\mu q^i)} \eta^i + \frac{\partial \La}{\partial (\partial_\mu q^i)} \partial_\nu q^i \, \xi^\nu - \La \xi^\mu + \mathfrak{g}^\mu.
   \label{consquant}
\end{equation}
In Eqs.~\eqref{noethidentity} and \eqref{consquant}, $\La$ is the Lagrangian density depending on the spacetime coordinates $x^\mu$, on the generalized variables $q^i$ and their derivatives $\partial_\mu q^i$, namely $\La \equiv \La(q^i,\partial_\mu q^i, x^\mu)$, while $\mathfrak{g}^{\mu}=\mathfrak{g}^{\mu}(x^\mu,q^i)$ are gauge functions. 

Considering the $f(R)$ gravity Lagrangian, let us search for spherically symmetric solutions, whose line element, expressed in geometric units and spherical coordinates $x^\mu = (t,r,\theta,\varphi)$, reads  
\begin{equation} \label{eq:general_metric}
\dd s^2 = -e^{\nu(r)} \dd t^2 + e^{\lambda(r)} \dd r^2 + \mathcal{M}(r) \dd \Omega^2,
\end{equation}
where $\dd \Omega^2\equiv \dd\theta^2+\sin^2\theta\dd\varphi^2$, $\mathcal{M}(r) d\Omega^2$ is a general 2-sphere, while $\nu(r)$ and $\lambda(r)$ are unknown smooth functions. 
In particular, we consider the case where the Ricci scalar depends only on the radius, \emph{i.e.} $R=R(r)$, and the  Birkhoff theorem holds. These assumptions thus   lead  to the static metric \eqref{eq:general_metric} (see  \cite{Capozziello:2010zz} for details). 

Therefore, in $f(R)$ gravity, the metric \eqref{eq:general_metric} leads to the point-like Lagrangian  \cite{Bajardi:2020osh, Capozziello:2007wc}
\begin{align} \label{lagrsphsym2}
&\Lagr = e^{\frac{\nu-\lambda}{2}} \frac{\mathcal{M}'^2}{2\mathcal{M}} f_R - e^{\frac{\nu-\lambda}{2}} \nu' \mathcal{M}' f_R - e^{\frac{\nu-\lambda}{2}}\mathcal{M}  \nu' R' f_{RR} \notag
\\
&- 2 e^{\frac{\nu-\lambda}{2}} R' \mathcal{M}' f_{RR} + e^{\frac{\nu+\lambda}{2}}\left[(\mathcal{M} R-2) f_{R}-\mathcal{M} f\right],
\end{align}
where the prime denotes the derivative with respect to the radial coordinate $r$. Notice that, though the configuration space is four-dimensional, the tangent space ${\cal{TQ}}$ contains only 7 variables:
\begin{equation}
\begin{cases}
\displaystyle {\cal{Q}} &= \{\nu, \lambda, \mathcal{M}, R \},
\\
\displaystyle {\cal{TQ}} &= \{\nu, \lambda, \mathcal{M}, R, \nu',\mathcal{M}',R' \}.
\end{cases}
\label{minisuper}
\end{equation}
Consequently, from the Euler-Lagrange equation with respect to $\nu$, we obtain the following relation:
\begin{align}
e^\lambda=& \dfrac{1}{2  \mathcal{M}\left[(2+\mathcal{M} R) f_{R}-\mathcal{M} f\right]} \Big\{ 2 \mathcal{M}^{2} f_{R R} \nu' R'  \nonumber
\\
&+2 \mathcal{M} f_{R} \nu' \mathcal{M}^{\prime}+4 \mathcal{M} f_{R R} \mathcal{M}^{\prime} R^{\prime}+ f_{R} \mathcal{M}^{\prime 2} \Big\}.
\label{EL00}
\end{align}
The other two Euler-Lagrange equations, calculated with respect to $\lambda$ and $\mathcal{M}$, provide the two remaining field equations:
\begin{subequations}
\begin{align}
     & 2 \mathcal{M} \left[f_R \left(-2 \mathcal{M}''+\mathcal{M}' \lambda'+2 e^{\lambda}\right)-2 \mathcal{M}' R' f_{RR}\right]\notag
     \\
     & +2 \mathcal{M}^2 \left(-2 f_{RRR} R'^2-2 R'' f_{RR} +\lambda' R' f_{RR} + e^{\lambda} R f_R\right)\notag
     \\
    &   -\mathcal{M}'^2 f_R -2 \mathcal{M}^2 e^{\lambda} f = 0,
   \\
   \notag\\
  & -\mathcal{M}'^2 f_R + \mathcal{M} \left\{2 \mathcal{M}' R' f_{RR} + f_R \left[2 \mathcal{M}''+\mathcal{M}' \left(\nu'-\lambda'\right)\right]\right\} \notag
  \\
  & +\mathcal{M}^2 \left\{f' \left(\lambda' \nu'-2 \nu''-\nu'^2\right)+2 e^{\lambda} R f_R \right.\notag
  \\
  & \left. -2 \left[2 f_{RRR} R'^2+2 R'' f_{RR} +R' f_{RR} \left(\nu'-\lambda'\right)\right]\right\} \notag
  \\
  &-2 \mathcal{M}^2 e^{\lambda} f = 0.
\end{align}
\end{subequations}
It is worth pointing out that the introduction of the function $\mathcal{M}(r)$ allows obtaining a non-dissipative Lagrangian so that the energy condition can be used to recover the $tt$ component of the Einstein field equations, namely the Euler-Lagrange equation with respect to $\nu$. More precisely, the energy condition is a requirement of zero energy that can be applied to non-dissipative Lagrangians in order to make the Lagrange multiplier method equivalent to the variational approach. It generally reads

\begin{equation}
    E_\La = \dot{q}^i \frac{\partial \La}{\partial \dot{q}^i} - \La = 0,
    \label{ECond}
\end{equation}
with $q^i$ being variables in the configuration space and with the \qm{dot} denoting the derivative with respect to an affine parameter. 

The Noether symmetry existence condition yields a system of 6 partial differential equations, which can be reduced after neglecting linear combinations.  In our case, the Noether theorem provides a unique solution for the function $f$, namely $f(R) = f_0 R^k$ with $k\in\mathbb{R}$ \cite{Capozziello:2007wc, FarasatShamir:2012qh, Capozziello:2009jg}. As a consequence, the symmetry generator $\mathcal{X}$ and the conserved quantity $j$ turn out to be
\begin{subequations}
\begin{align}
\displaystyle {\cal{X}} &= \alpha_0(3-2k) e^\nu \partial_\nu - \alpha_0 \mathcal{M} \partial_\mathcal{M} + \alpha_0 R \partial_R,
\\ 
\displaystyle j &= 2k \alpha_0 \mathcal{M} R^{2 k-3}[2 k+(k-1) \mathcal{M} R] \nonumber \\
&\ \ \times \displaystyle \left[(k-2) R \nu'-\left(2 k^{2}-3 k+1\right) R'\right] e^\nu.
\end{align}
\label{firstsetofsolf(R)}
\end{subequations}
A Schwarzschild-de Sitter solution is recovered for constant $R$. It is worth noticing that, for power-law models as $f(R)\sim R^k$,  dynamics can be reduced and solved both at cosmological \cite{Capozziello:2002rd,Capozziello:2003gx, Lambiase:2006dq, Nojiri:2017ncd} and astrophysical \cite{Clifton:2005aj,Martins:2007uf,Capozziello:2019klx} scales. 

\subsection{Black hole solutions}
\label{sec:SS_metrics}
The integral of motion $j$ in Eq.~\eqref{firstsetofsolf(R)} can be adopted to determine two relations among $\nu$, $\mathcal{M}$, and $R$, which can be distinguished for $k\neq 2$,
\begin{align}
&e^\nu = f_0 R^{\frac{2 k^{2}-3 k+1}{k-2}}
\notag\\
&\times \left\{1+  j \int \frac{R^{\frac{4 k^{2}-9 k+5}{2-k}} d r}{2 \alpha_0 k(k-2) \mathcal{M}[2 k+(k-1) \mathcal{M} R]}\right\},
\end{align}
and for $k=2$,
\begin{equation}
e^\nu =-\frac{j}{12 \alpha_0 r^{2}\left(4+r^{2} R\right) R R^{\prime}}.
\end{equation}
Specific subcases of the above general solutions are discussed in Refs.~\cite{Capozziello:2007wc,Clifton:2005aj,Clifton:2006ug,Bajardi:2022ocw}, where the first-order approximation of the field equations around $k = 1$ is taken into account. Hence, setting $k = 1+ \epsilon$ with $|\epsilon| \ll 1$, the metric components can be written as
\begin{subequations}
\begin{align}
e^\nu &=r^{\frac{2 \epsilon(1+2 \epsilon)}{1-\epsilon}}+\frac{C_{1}}{r^{\frac{1-4 \epsilon}{1-\epsilon}}}, \label{eq:e^nu} \\
e^\lambda &=\left\{ \left[\frac{(1-\epsilon)^{2}}{\left(1-2 \epsilon+4 \epsilon^{2}\right)[1-2 \epsilon(1+\epsilon)]}\right]\right.\notag\\
&\left.\qquad\times\left(1+\frac{C_{1}}{r^{\frac{1-2 \epsilon+4 \epsilon^{2}}{1-\epsilon}}}\right)\right\}^{-1}.
\label{eq:e^lambda}
\end{align}
\end{subequations}
Let us consider the generic static and spherically symmetric geometry \eqref{eq:general_metric}, where we choose $\mathcal{M}(r)=r^2$, corresponding to standard 2-spheres. This class of metrics has been extensively studied in Refs.~\cite{Capozziello:2007wc, Clifton:2005aj, Clifton:2006ug}. 

In order to select a specific BH solution, we need to determine the parameters $\epsilon$ and $C_1$ such that Eq.~\eqref{eq:general_metric} verifies the BH features, namely: $(i)$ the existence of a physical singularity, $(ii)$ the presence of an event horizon, $(iii)$ the asymptotic flatness. 

To avoid divergences when $r \rightarrow \infty$, the exponents of $r$ in Eqs.~\eqref{eq:e^nu} and \eqref{eq:e^lambda} give rise to the following system of inequalities
\begin{align}
&\frac{2 \epsilon  (2 \epsilon +1)}{1-\epsilon }\leq 0,\ \ \frac{1-4 \epsilon }{1-\epsilon }\geq 0,\ \
\frac{4 \epsilon ^2-2 \epsilon +1}{1-\epsilon }\geq 0,
\end{align}
from which we eventually obtain
\begin{align} \label{eq:eps_range}
-\frac{1}{2}\leq \epsilon \leq 0.    
\end{align}
The case $\epsilon=0$ corresponds to the Schwarzschild geometry. Taking into account the interval \eqref{eq:eps_range}, we can choose $\epsilon=-0.01$, for the purpose to investigate small departures from the Schwarzschild solution. 
Moreover, we can set $C_1=-r_{\rm 0}^{1.01}$, where $r_{\rm 0}$ is a real constant  with dimension of length. Then, the metric components \eqref{eq:e^nu} and \eqref{eq:e^lambda} become
\begin{subequations} \label{eq:metric2}
\begin{align}
e^{\nu(r)}&=\frac{1}{r^{0.02}}-\frac{r_0^{1.01}}{r^{1.03}}\,, \qquad e^{\lambda(r)}=\frac{1.02}{1-\left(\frac{r_0}{r}\right)^{1.01}}.
\label{eq:elambda_eps}  
\end{align}
\end{subequations}
The parameter $r_{\rm 0}$ may be determined by imposing the photonsphere to be located at $r_{\rm ps}=3.00M$, as in the Schwarzschild case, with $M$ being the BH mass. In this case, we ensure our solution mimics the Schwarzschild properties with the following benefits:  $(i)$ computations can be easily performed; $(ii)$ physically viable results can be provided. Considering the null geodesic equation, we obtain (see Ref.~\cite{Defalco2021}, for details)
\begin{equation} \label{eq:null_geodesic}
E_p^2=V_p(r)\equiv\frac{e^{\nu(r)} L_p^2}{r^2}, 
\end{equation}
where $E_p$ and $L_p$ are the conserved energy and angular momentum along the photon trajectory, respectively. 
Imposing the condition for the orbit stability, $\frac{{\rm d}V_p(r)}{{\rm d}r}=0$, where $V_p(r)$ is the photon geodesic potential and substituting $r=r_{\rm ps}$  into Eq.~\eqref{eq:null_geodesic}, we find $r_{\rm 0}=2.01M$. Therefore,  we get
\begin{equation}
\label{eq:BH_sol}
e^{\nu(r)}=\frac{1}{r^{0.02}}-\frac{2.02}{r^{1.03}}\,,\qquad
e^{\lambda(r)}=\frac{1.02}{1-\frac{2.02}{r^{1.01}}}\,, 
\end{equation}
where we have set $M=1$ for simplicity.
In this case, the distance between the BH and the observer is identified by the parameter $r_{\rm obs}$, which we choose to be $r_{\rm obs}=10^{10}M$ as the distance to Sgr A$^\star$, the BH at the center of our Galaxy\footnote{The chosen value of $r_{\rm obs}$ manifests how the metric \eqref{eq:BH_sol} behaves at infinity. However, it can be shown that different higher values of $r_{\rm obs}$ provide results comparable to those presented here.}\cite{Akiyama2022image}, which gives 
\begin{equation} \label{eq:BH_metric}
\lim_{r\to r_{\rm obs}}e^{\nu(r)}=0.64,\qquad  \lim_{r\to r_{\rm obs}}e^{\lambda(r)}=1.02.
\end{equation}
The BH event horizon $r_{\rm h}$ is obtained by $g_{tt}=0$ and $1/g_{rr}=0$, which leads to $r_{\rm h}\equiv r_0=2.01M$. The metric \eqref{eq:BH_sol} presents also a physical singularity located at $r=0$.

\section{Probing the BH solution in presence of gravity only}
\label{sec:analysis_gravity}
Let us  investigate the above static and spherically symmetric BH solution \eqref{eq:BH_sol} via the comparison with the Schwarzschild spacetime. In both cases, gravity is the only acting force. The astrophysical techniques we are going to take into account are the following:
\begin{enumerate}[label=\Alph*.]
\item the ray-tracing method for reconstructing the BH image in the observer screen;
\item the epicyclic frequencies;
\item the BH shadow profile. 
\end{enumerate}
\begin{figure*}[ht!]
    \centering
    \includegraphics[scale=0.42]{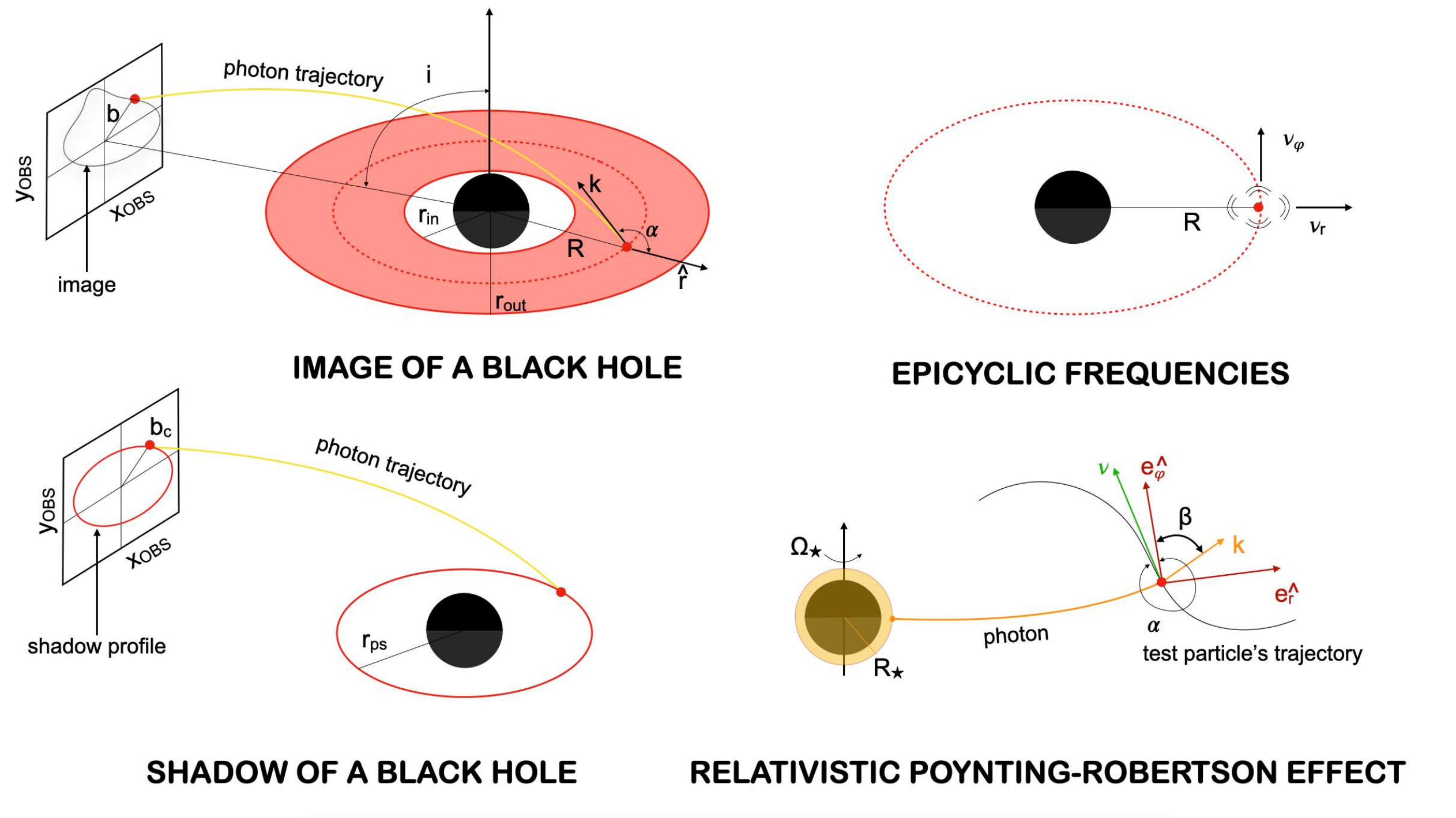}
    \caption{Sketch of different astrophysical techniques exploited to detect departures from the Schwarzschild geometry. They can be divided in two astrophysical situations: (1) only gravity is acting on the surrounding matter outside the BH event horizon; (2)  also X-ray radiation processes are involved, besides gravity. The first case includes the image of a BH, the epicyclic frequencies, and the BH shadow profile, whereas the second one accounts for the relativistic PR effect. }
    \label{fig:Fig_astro}
\end{figure*}

\subsection{Black hole imaging}
\label{sec:BH_image}
Let us first consider the photon four-momentum $k^\mu$ in the equatorial plane:
\begin{equation}
\boldsymbol{k}:=\left[-E_p,\sqrt{\frac{1}{e^{\lambda(r)}}\left(\frac{E_p^2}{e^{\nu(r)}}-\frac{L_p^2}{r^2}\right)},L_p\right].    
\end{equation}
Defining $b:=L_p/E_p$, as the photon impact parameter, we introduce the light bending angle $\psi(r,b)$ as \cite{Defalco2021}
\begin{align} \label{eq:LB}
\psi(r,b)&:=\int_\mathcal{R}^\infty \frac{\tilde{k}^\varphi}{\tilde{k}^r}{\rm d}r\notag\\
&=\int_\mathcal{R}^\infty\frac{b}{r^2}\left[\frac{1}{e^{\lambda(r)}}\left(\frac{1}{e^{\nu(r)}}-\frac{b^2}{r^2}\right)\right]^{-1/2}{\rm d}r,    
\end{align}
where $\tilde{k}^\mu=k^\mu/E_p$ and $\mathcal{R}$ is the emission radius.  The above formula describes how a photon geodesic bends along the path connecting the emission region, nearby the BH, with the observer, located at a very large distance from the gravitational source. Therefore, it is possible to ray-trace the BH environment with respect to a static observer (SO) located at distance $r_{\rm obs}$ and inclined with respect to the axis orthogonal to the BH equatorial plane of an angle $i$\footnote{It is worth noticing that  the closer the observer to the BH equatorial plane, the stronger the relativistic effects.} \cite{Luminet1979,Defalco2016}. To do this, we introduce the emission angle $\alpha$, which corresponds to the angle between the photon four-momentum vector $\boldsymbol{k}$ and the radial direction $\hat {\boldsymbol{r}}$ on the equatorial plane. 

It is possible to relate the photon impact parameter to the photon emission angle $\alpha$ and radius $\mathcal{R}$ via
\begin{equation}
b=\frac{\mathcal{R}\sin\alpha}{\sqrt{e^{\nu(\mathcal{R})}}}.    
\end{equation}
We also define the critical photon impact parameter as
\begin{equation} \label{eq:b_c}
b_c=\frac{r_{\rm ps}}{\sqrt{e^{\nu(r_{\rm ps})}}}=5.25M\,.
\end{equation}
As such, for $b>b_c$, the light rays reach the observer, whereas, for $b\le b_c$, the photons are swallowed by the BH. 
Another useful quantity is the maximum emission angle $\alpha_{\rm max}$, defined (for $\mathcal{R}>r_{\rm ps}$) as
\begin{equation}
\alpha_{\rm max}=\pi-\arccos\left[b_c\frac{\sqrt{e^{\nu(\mathcal{R})}}}{\mathcal{R}}\right].    
\end{equation}
It is worth noticing that Eq.~\eqref{eq:LB} is defined for $\alpha\in[0,\alpha_p=\pi/2]$. On the other hand, for $\alpha\in[\alpha_p,\alpha_{\rm max}]$, turning points occur and a symmetrization process is needed. Thus, we first consider $\alpha_S=\pi-\alpha$ and apply Eq.~\eqref{eq:LB} to calculate $\psi_S$. To correctly associate the right bending angle $\psi$, we then compute the periastron $p$, which is the minimum distance of the light ray from the BH, whose value is obtained by solving the following algebraic equation \cite{Defalco2021}: 
\begin{equation}
p^2-b^2e^{\nu(p)}=0,    
\end{equation}
which depends on $b$ and therefore on the emission angle $\alpha$. We then calculate the periastron bending angle $\psi_p$ as
\begin{equation} \label{eq:LBp}
\psi_p(p,b_p)=\int_p^\infty\frac{b_p}{r^2}\left[\frac{1}{e^{\lambda(r)}}\left(\frac{1}{e^{\nu(r)}}-\frac{b_p^2}{r^2}\right)\right]^{-1/2}{\rm d}r,    
\end{equation}
where $b_p=p/\sqrt{e^{\nu(p)}}$. Finally, the light bending angle $\psi$, related to $\alpha$, is given by $\psi=2\psi_p-\psi_S$ \cite{Defalco2016}. It is worth observing that, in our case, the upper bounds of integration in Eqs.~\eqref{eq:LB} and \eqref{eq:LBp} are replaced by $r_{\rm obs}$.

In order to reconstruct the image of matter moving around a BH in the observer plane, we adopt the celestial coordinates $(x_{\rm obs},y_{\rm obs})$, defined as 
\begin{equation}
\begin{cases}
x_{\rm obs}&=-b\,\dfrac{\sin\varphi}{\sin\psi},\\
y_{\rm obs}&=-b\,\dfrac{\cos i \cos\varphi}{\sin\varphi},
\end{cases}    
\end{equation}
where $\varphi\in[0,2\pi]$ is the azimuthal angle. For each $\varphi$, we associate the related $\psi$ using $\cos\psi=\sin i\cos\varphi$ and finally we calculate $\alpha$ via interpolation.

We consider the presence of an accretion disk around the BH and  assume, for the sake of simplicity, that matter moves on Keplerian orbits at different radii. Therefore, the matter velocity in the disk frame is given by
\begin{equation}
U^{\hat\alpha}=\left(1,0,\Omega_K \frac{\sin i\sin\varphi}{\sin\varphi}\right),    
\end{equation}
where 
\begin{equation} \label{eq:omega_k}
\Omega_K:=\sqrt{\frac{e^{\nu(\mathcal{R})}\nu'(\mathcal{R})}{2\mathcal{R}}}=\frac{\sqrt{\frac{1.04}{\mathcal{R}^{1.03}}-\frac{0.01}{\mathcal{R}^{0.02}}}}{\mathcal{R}}   
\end{equation}
is the Keplerian angular velocity and the prime denotes the derivative with respect to $\mathcal{R}$. In our case, the innermost stable circular orbit (ISCO) radius associated to test particles is $r_{\rm isco}=6.23M$\footnote{We refer to Ref.~\cite{Defalco2021} for details, or to Sect.~\ref{sec:BH_epi_freq} for an alternative derivation of the ISCO radius.}.

We then define the gravitational redshift through the relation \cite{Defalco2016,Defalco2021}
\begin{equation}
(1+z)^{-1}:=\frac{\sqrt{e^{\nu(r)}-\Omega_K^2\mathcal{R}^2}}{1+b\Omega_K\frac{\sin i\sin\varphi}{\sin\varphi}}.
\end{equation}
We consider an accretion disk around a BH, which extends from $r_{\rm in}\equiv r_{\rm isco}=6.23M$ to $r_{\rm out}=100M$. In Fig.~\ref{fig:Fig_astro}, we provide a visual scheme of what described above.
\begin{figure}[ht!]
    \centering
    \vbox{
    \includegraphics[scale=0.3]{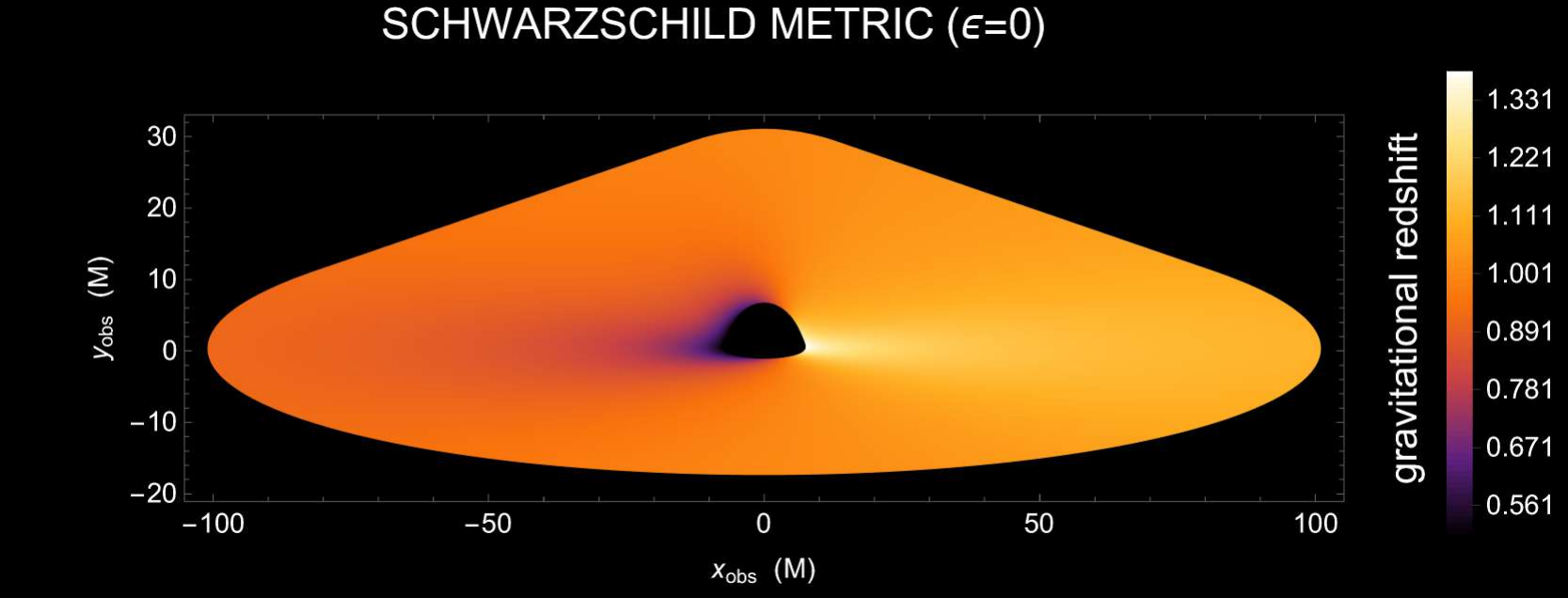}
    \includegraphics[scale=0.3]{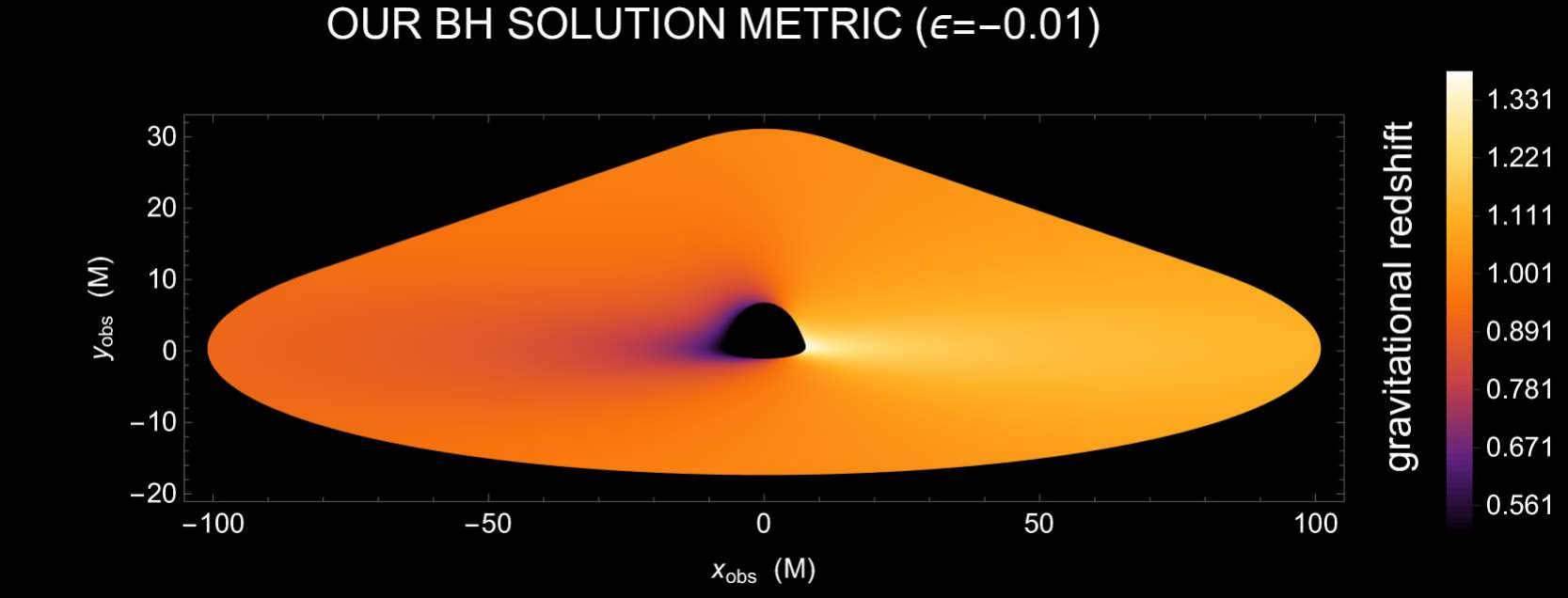}}
    \caption{Accretion disk dynamics around a BH in coordinates $(x_{\rm obs},y_{\rm obs})$ for a distant observer with an inclination angle $i=80^\circ$. The disk extends from $r_{\rm in}=6.23M$ to $r_{\rm out}=100M$. Matter moves with Keplerian angular velocity $\Omega_K$. For each point, we report the gravitational redshift (see lateral bars).}
    \label{fig:Fig1}
\end{figure}

\begin{figure*}[ht!]
    \centering
    \hbox{
    \includegraphics[scale=0.29]{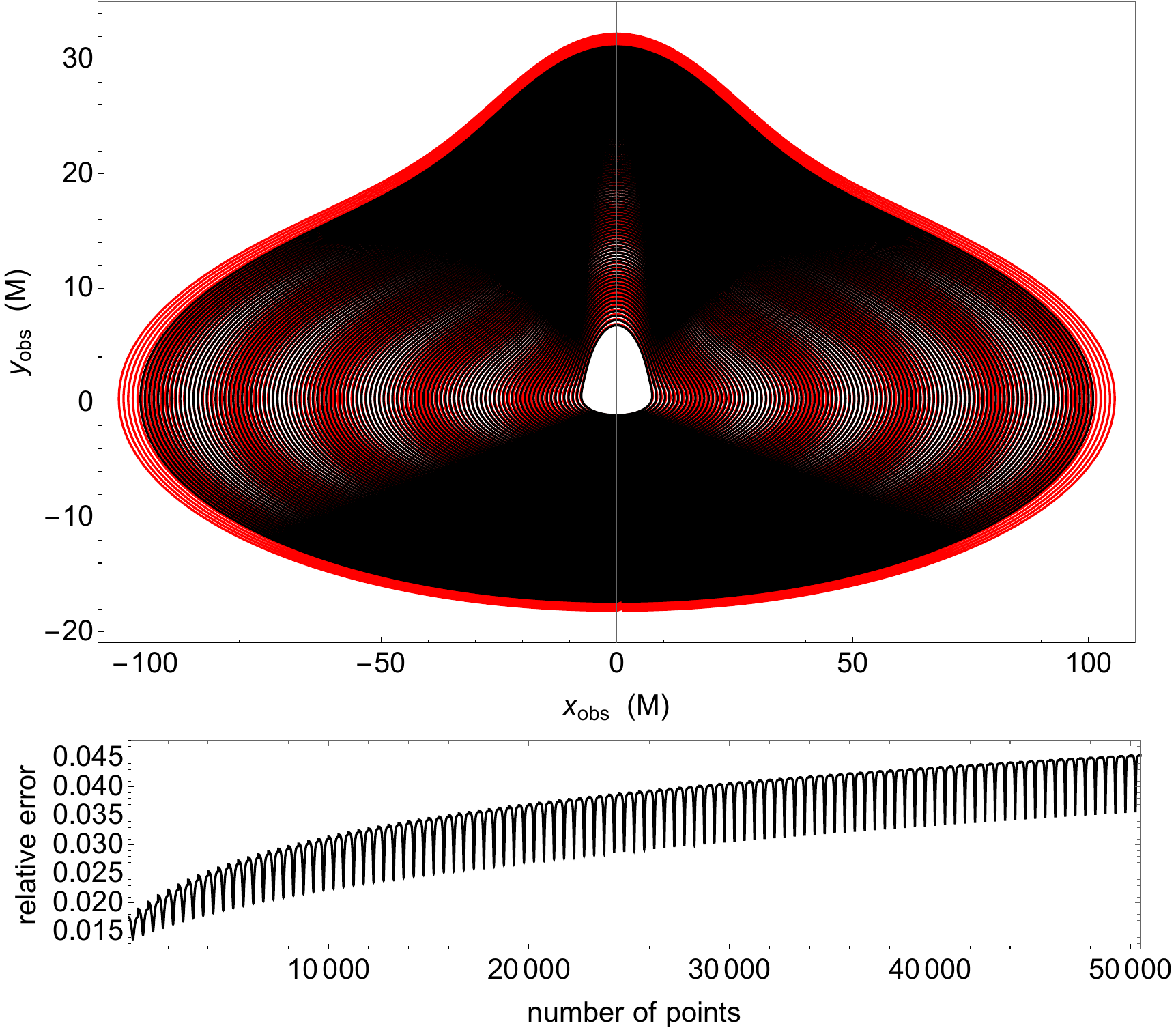}\hspace{0.3cm}
    \includegraphics[scale=0.29]{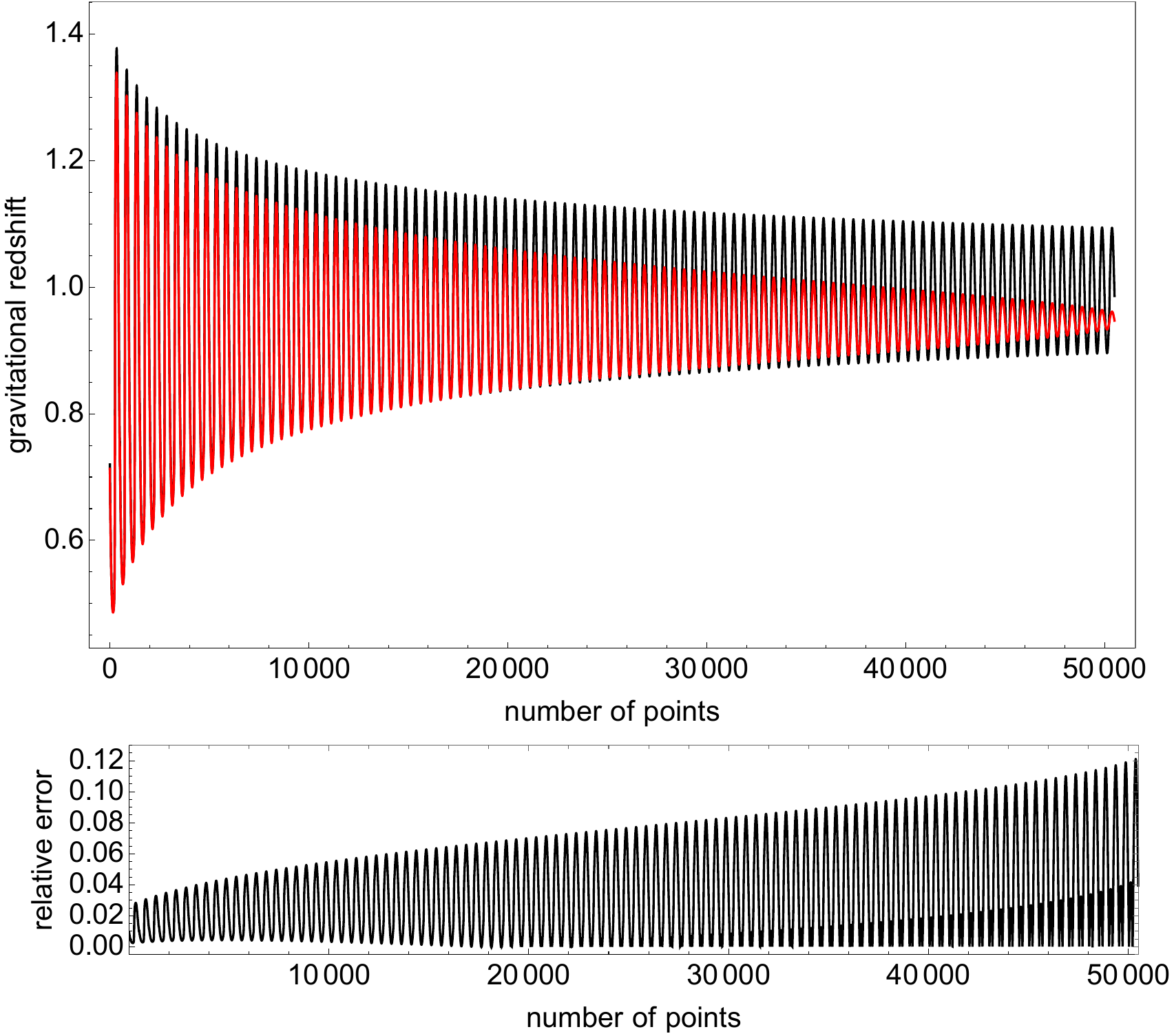}}
    \caption{In the upper plots, black lines refer to the Schwarzschild metric whereas red ones to our BH solution. \emph{Left upper panel.} The two overlapped disks are individually composed of 101 rings, each one consisting of 500 points, for a total of 50500 points. \emph{Left lower panel.} Relative errors between the corresponding disk-points in the two metrics. \emph{Right upper panel}. Gravitational redshift \emph{versus} the disk-points. \emph{Right lower panel.} Relative errors related to the gravitational redshift.}
    \label{fig:Fig1bis}
\end{figure*}

In Fig.~\ref{fig:Fig1}, we show the BH image related to the Schwarzschild ($\epsilon =0$) and our BH ($\epsilon =-0.01$) solution. We note that the two images appear as identical at the first sight. In fact, a more detailed analysis must be performed in order to spot the differences between the two spacetimes not only in terms of geometric structure, but also through the gravitational redshift values. To this end, we develop  a further investigation in Fig.~\ref{fig:Fig1bis}. We first study the shape of the disk, which is closely related to the background spacetime geometry. In such a way, we can visually distinguish between the regions, where departures from the Schwarzschild metric are present. The simulations have been performed using the same number of points and the same radii to shape the whole disk. Therefore, to quantify deviations from the Schwarzschild geometry, we can calculate the relative error on each disk-point\footnote{The relative error used in our calculations is defined as $\left|\frac{f_{\rm Sch}-f_{\rm sol}}{f_{\rm Sch}}\right|$, where $f_{\rm Sch}$ and $f_{\rm sol}$ are quantities related to the Schwarzschild and our $f(R)$ BH solution, respectively.}. From this analysis, we can see that strong departures can be detected moving closer to $r_{\rm out}$. This is due to the fact that our BH solution does not properly go to infinity. The minimum relative error, $\sim1.5\%$, occurs at the ISCO radius, whereas the maximum, $\sim4.5\%$, is in correspondence of $r_{\rm out}$. The analysis of the gravitational redshift leads to the same conclusions, since the relative error increases when approaching the outer edge of the disk. The outcomes of our analysis can be compared to the Event Horizon Telescope (EHT) data, whose angular resolution is $(\lambda/D)\sim (20-25)\ \mu\mbox{as}$ \cite{EHT2022}. Indeed, if we consider, for example, the Sgr A$^\star$ source with mass $M\sim 4\times10^6\ M_\odot$ and distance to the Earth $d\sim 8\ {\rm kpc}$, we find that the ISCO radius is  $30.16\ \mu\mbox{as}$, being therefore in the field of view of EHT. However, the aforementioned departures from GR cannot be currently appreciated \cite{Akiyama2022} as shorter wavelengths would be needed. However, these upgrades may be attained with one of the next releases.

In this astrophysical context, the observational X-ray data, consisting of the persistent flux (or equivalently luminosity) emitted by the accretion disk, are provided by the present INTEGRAL, XMM-Newton, Swift instruments \cite{Falanga2015}, and by near-future space missions, like Athena \cite{Barret2016} and e-XTP \cite{Zhang2018}. Modulo the emission properties of the disk, the flux generally depends on the gravitational redshift and the solid angle, which takes into account the light bending \cite{Defalco2016}. In our approach, we separately examine the fundamental contributions (light bending and gravitational redshift) that, generally, should be inferred from the data reduction and analysis.

Alternative approaches aim to fit observational data (related to almost static or very slowly rotating compact objects) by making use of the Schwarzschild metric. If any deviations is detected, other metrics may be employed in the analysis \cite{Sheoran2018,Abdikamalov2020,Tripathi2021,Tripathi2022,Zhang2022} along with suitable reconstruction methodologies \cite{Volkel2020,Defalco2021wh,Defalco2021EF,Lara2021}. Further alternative strategies involve the use of general BH parametrizations to be fitted with data, in order to trace back the underlying geometry \cite{Glampedakis2006,Johannsen2010,Rezzolla2014,Konoplya2016,Medeiros2020}. 

\subsection{Epicyclic frequencies}
\label{sec:BH_epi_freq}
\begin{figure*}
    \centering
    \hbox{
    \includegraphics[scale=0.3]{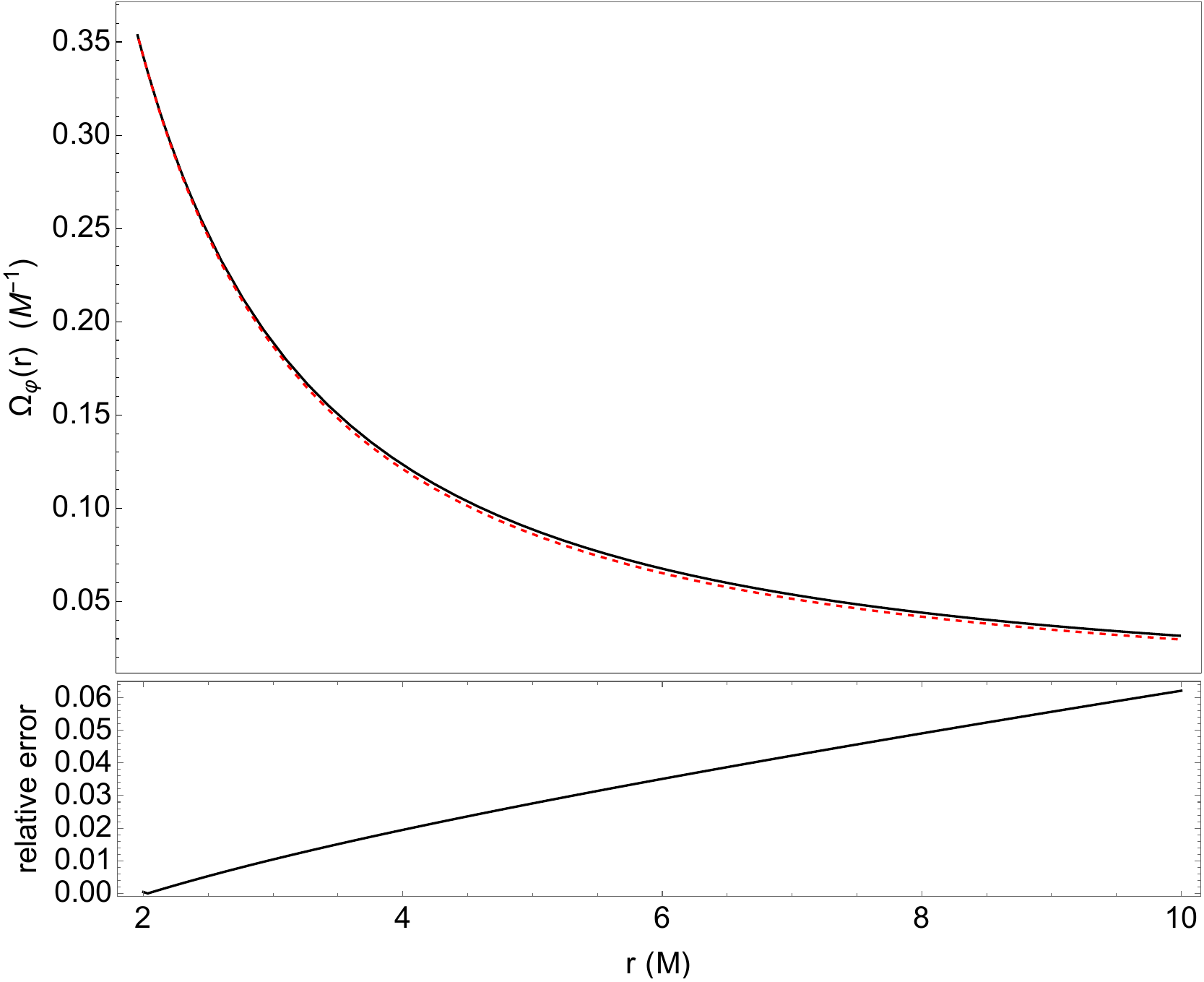}\hspace{0.5cm}
        \includegraphics[scale=0.3]{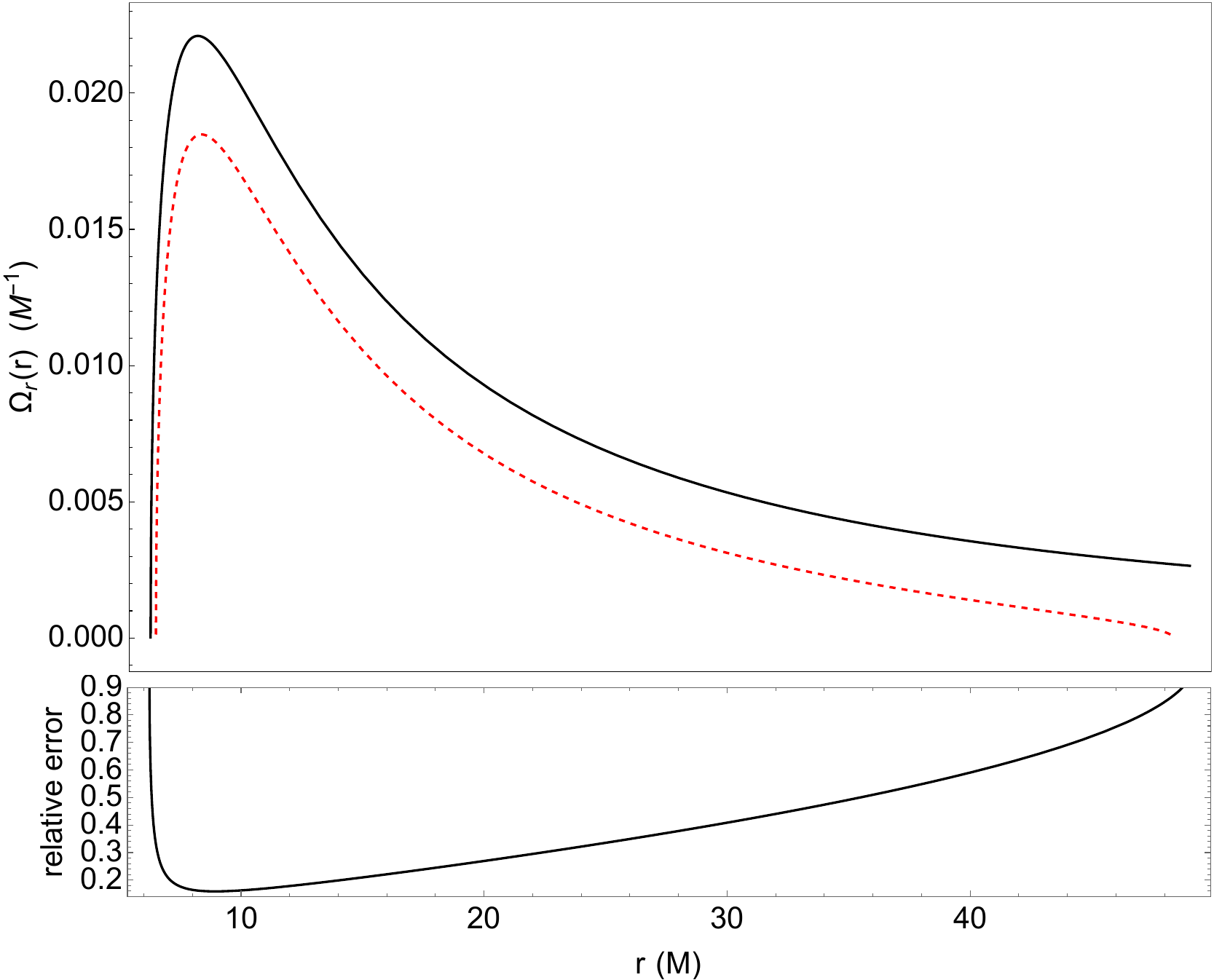}}
    \caption{Epicyclic angular frequencies $\Omega_\varphi$ (left panel) and $\Omega_r$ (right panel). The solid black lines refer to the Schwarzschild spacetime, whereas the dashed red lines to our BH solution.  The relative errors are reported in the bottom panels.}
    \label{fig:Fig2}
\end{figure*}

Another approach for investigating compact objects is represented by the epicyclic frequencies. These generally encode the gravitational perturbations on the classical circular motion along the three fundamental directions $\nu_r$, $\nu_\theta$, and $\nu_\varphi$. However, in a spherically symmetric spacetime, we have only $\nu_r$ and $\nu_\varphi$, since $\nu_\theta\equiv\nu_\varphi$ (see Fig.~\ref{fig:Fig_astro}). These quantities exhibit a series of advantageous properties \cite{Ingram2019,Defalco2021}, such as: strong relation with the background spacetime geometry; production of observational effects in extreme gravitational field regimes; direct measurement via quasi-periodic oscillations (QPOs) through the present (\emph{e.g.}, XMM-Newton \cite{Falanga2015}, LOFT \cite{Feroci2016}) and near future (\emph{e.g.}, e-XTP \cite{Zhang2018}, IXPE \cite{Soffitta2013}) observations; great availability of already existing observational data, extracted by applying the most suited QPO models to the gravitational system under investigation. 

A common procedure makes use of the related epicyclic angular velocities $\left\{\Omega_r,\Omega_\varphi\right\}$, whose expressions are \cite{Defalco2021EF}
\begin{subequations}
\begin{align}
\Omega_\varphi&:=\sqrt{\frac{-g_{tt}'(r)}{2r}}=\sqrt{\frac{e^{\nu(r)}\nu'(r)}{2r}},\\
\Omega_r&:=\sqrt{\frac{g_{tt}^2(g^{tt})''+6\Omega_\varphi ^2}{2g_{rr}}}\notag\\
&=e^{\nu (r)}\sqrt{\frac{e^{-\nu (r)} \left[r \nu ''(r)-r \nu '(r)^2+3 \nu '(r)\right]}{2re^{\lambda (r)}}},
\end{align}
\end{subequations}
where $\Omega_\varphi\equiv\Omega_k$ (cf. Eq.~\eqref{eq:omega_k}). It is possible to calculate $r_{\rm isco}$ by determining the $r$-value such that $\Omega_r$ is zero, in agreement with the calculations performed by resorting to the timelike geodesic equations \cite{Defalco2021}.

In Fig.~\ref{fig:Fig2}, we display the two epicyclic angular frequencies both for the Schwarzschild case (black solid lines) and for our BH solution (red dashed lines), with the related relative errors. We can clearly notice a relevant difference along the radial direction, more precisely close to the ISCO radius up to $r\sim7M$, and also moving further from the gravitational source. Furthermore, the radial epicyclic frequency is defined up to $r\approx49.31M$, where it becomes zero. After this value, the radius becomes imaginary. The error bands of our analysis are within the current observational sensitivity (see \emph{e.g.}, Ref. \cite{Motta2013}, for details).

\subsection{Shadow profile}
\label{sec:BH_shadow}
The last method we focus on takes into account the BH shadow profile. The presence of a BH can be inferred from the interaction with the surrounding matter through the formation of accretion structures. The BH gravitational potential energy is converted into kinetic energy and matter increases its temperature due to dissipation effects occurring in the disk plasma, causing electromagnetic emission mainly in the X-ray energy band. The strong gravitational field affects the light bending, so it is possible to obtain a two-dimensional dark zone in the observer plane, known as BH shadow \cite{Falcke2000}, through the ray-tracing technique. This system is constructed by considering not only first-order direct photon images, but also higher-order terms, arising from photon trajectories undergoing one or more loops in the close vicinity of the BH \cite{Luminet1979,Falanga2021}. Therefore, this dark region encircled by a bright emission ring, only depends upon the BH spacetime geometry (see Fig.~\ref{fig:Fig_astro}). In general, the shadow profile of a non-rotating BH is a standard circle with radius $b_c$ \cite{Luminet1979}, whereas, for a rotating BH, the shadow will be elongated along the rotation direction. Regarding astrophysical BHs, their size and shape  closely depend on the BH mass and spin, as well as on the inclination of the observer with respect to the BH equatorial plane. Such a shadow is fundamental for extracting information on gravity in extreme field regimes.

\begin{figure*}
    \centering
    \hbox{
    \includegraphics[scale=0.3]{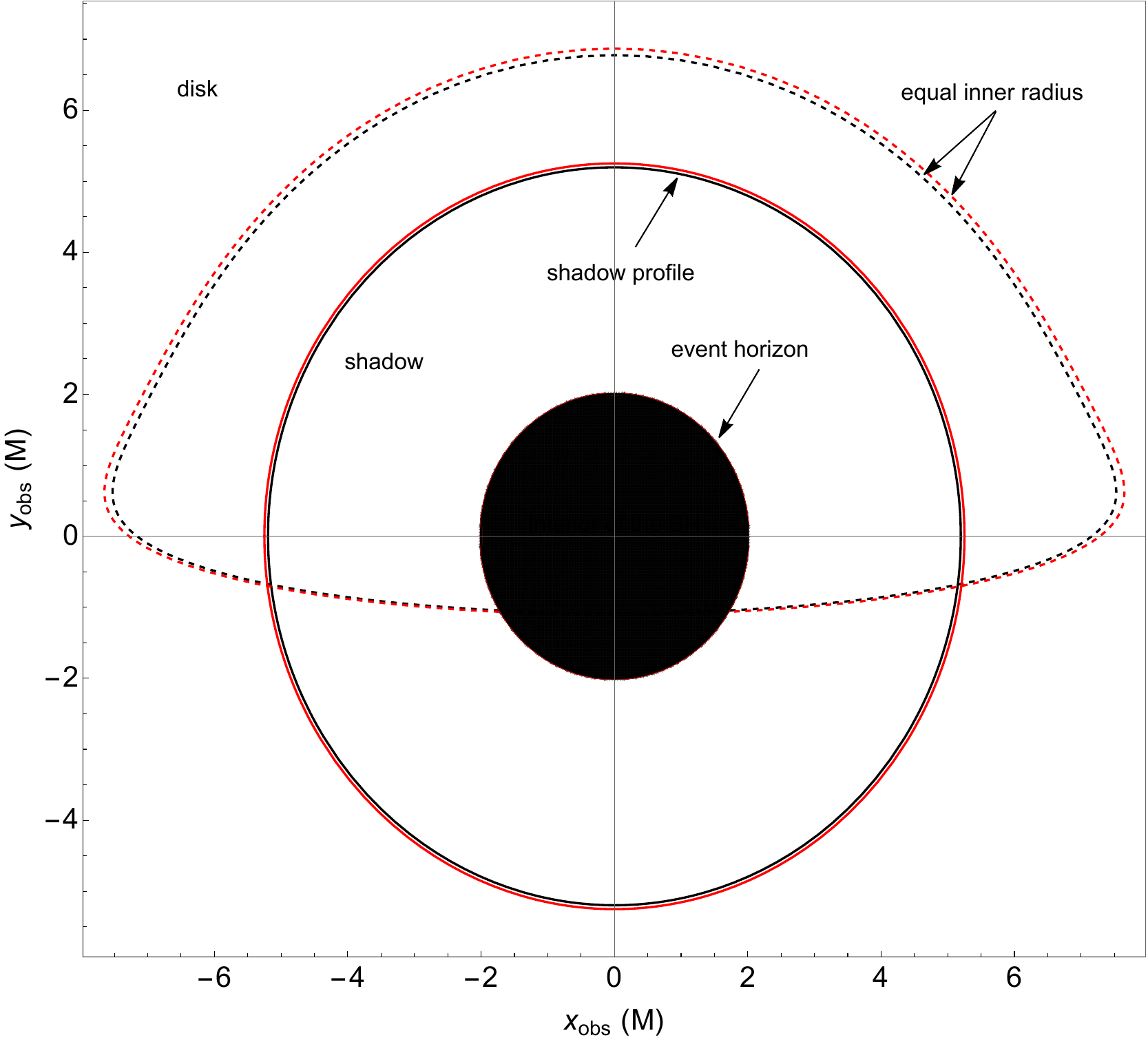}\hspace{0.5cm}
    \includegraphics[scale=0.3]{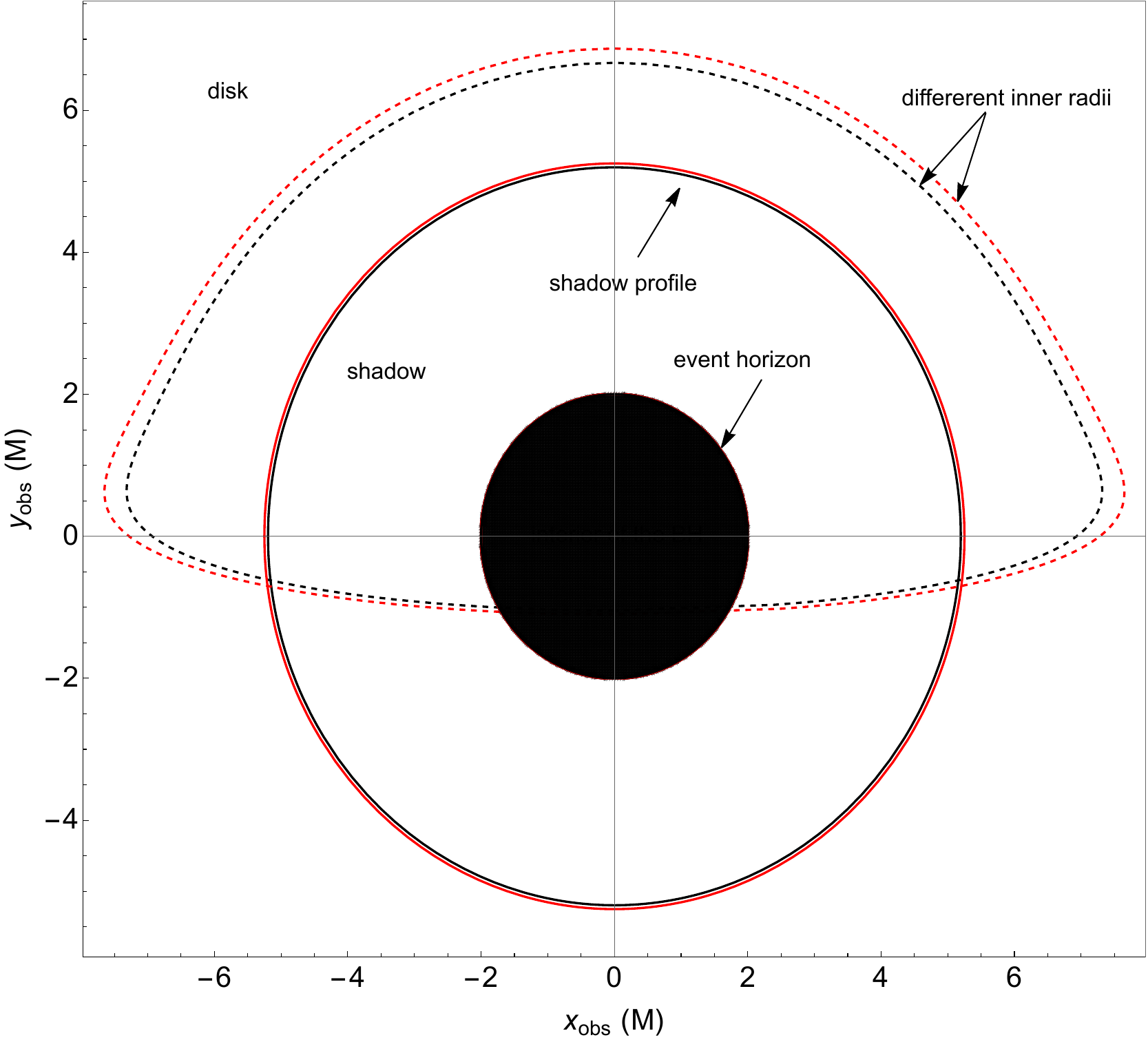}}
    \caption{ Inner disk radius, shadow profile, and BH event horizon as seen by a distant SO in its own coordinate frame $(x_{\rm obs},y_{\rm obs})$. The black and red lines refer to the Schwarzschild metric and our BH solution, respectively. \emph{Left panel.} The inner disk radius is set to the same value for both metrics, namely $r_{\rm in}=6.23M$, corresponding to the ISCO radius of our BH solution. \emph{Right panel.} The inner disk regions are set to the proper ISCO radii for both metrics.} 
    \label{fig:Fig3}
\end{figure*}

In Fig.~\ref{fig:Fig3}, we display the shadow profile of our BH solution (red lines) together with the Schwarzschild metric (black lines), as well as the ISCO radius and the BH event horizon. When the inner disk radius coincides with the ISCO, it is hard to spot differences both from the shadow and ISCO radius. However, if we consider the inner disk radius to be located at the proper ISCO value for each metric, some differences may emerge. Nevertheless, difficulties in detecting relevant departures may also occur in the latter case, when the observer inclination angle is $i\lesssim30^\circ$. Instead, discrepancies may be measured as the observer moves closer to the equatorial plane, due to the enhanced relativistic effects \cite{Luminet1979,Defalco2016}.

The observational data related to the shadow profile are essentially those provided by EHT. The possibility of detecting metric departures from this analysis follows the same discussions reported in Sect.~\ref{sec:BH_image}. Our case sets below the actual  wavelength of EHT, but near future upgrades will most likely reach the requested precision.

\subsection{Discussion of the  results}
\label{sec:discussion}
\renewcommand{\arraystretch}{1.4}
\begin{table*}[ht!]
    \caption{Summary of the results from our BH solution ($\epsilon=-0.01$) compared with the Schwarzschild metric's parameters.}
    \centering
    \scalebox{1.}{
    \begin{tabular}{|c|c|c|c|}
    \hline
    \quad{\rm\bf Parameter}$\quad$ &\quad {\rm\bf Units}$\quad$ &\quad {\rm\bf Our BH solution} $\boldsymbol{(\epsilon=-0.01)}\quad$ &\quad {\rm\bf Schwarschild} $\boldsymbol{(\epsilon=0)}\quad$  \\
    \hline
    $g_{tt}(r_{\rm obs})$ & -- & 0.64 & 1.00\\
    $g_{rr}(r_{\rm obs})$ & -- & 1.02 & 1.00\\
    $r_{\rm h}$ & $M$ & 2.01 & 2.00\\
    $r_{\rm ps}$ & $M$ & 3.00 & 3.00\\
    $r_{\rm isco}$ & $M$ & 6.23 & 6.00\\    
    \hline
    $\alpha_{\rm max}(r_{\rm isco})$ & rad & 136.85 & 168.76\\
    $\alpha_{\rm max}(r_{\rm out})$ & rad & 177.15 &179.42\\
    $\psi_{\rm max}(r_{\rm isco})$ & rad & 87.69 & 43.80\\
    $\psi_{\rm max}(r_{\rm out})$ & rad & 137.73 & 38.78\\
    $\psi_{\rm p}(r_{\rm isco})$ & rad & 70.31 &28.73\\
    $\psi_{\rm p}(r_{\rm out})$ & rad & 70.31 &19.68\\
    $b_{\rm c}$ & $M$  & 5.25  & 5.20\\
    $b(r_{\rm isco},\alpha_p)$ & $M$  & 7.68  &7.55\\
    $b(r_{\rm out},\alpha_p)$ & $M$  & 105.59  &101.02\\    
    \hline    
    $\Omega_{\rm K}(r_{\rm isco})$ & $M^{-1}$ & 0.06 &0.07\\    
    $\Omega_{\rm K}(r_{\rm out})$ & $M^{-1}$ & $10^{-4}$ & $10^{-3}$\\
    $\Omega_{\rm r}(r_{\rm isco})$ & $M^{-1}$ & 0.00 &0.00\\    
    $\Omega_{\rm r}(r_{\rm out})$ & $M^{-1}$ & 0.00\footnote{The radial epicyclic frequency is defined up to $r\approx49.31M$  (see Sect.~\ref{sec:BH_epi_freq}).} &$10^{-3}$\\    
    \hline    
    \end{tabular}}
    \label{tab:astrophysics}
\end{table*}
Let us now discuss results obtained so far. For a more quantitative comparison, in Table~\ref{tab:astrophysics}, we report the outcomes of our solution ($\epsilon=-0.01$) compared with the Schwarzschild metric ($\epsilon=0$).
The first significant difference relies on the value of the metric time component at $r_{\rm obs}$, as it can be clearly seen in Fig.~\ref{fig:Fig1bis}.
 
Furthermore, the ray-tracing process in both spacetimes could reveal metric departures, as one can immediately note from the values of $\alpha_{\rm max},\psi_{\rm max},\psi_p$, and $b$. Therefore, we can conclude that a careful analysis of the accretion disk image could show whether there are departures from the Schwarzschild geometry. Nevertheless, this strategy is mainly based on the time component of the metric $g_{tt}$, rather than $g_{rr}$. Indeed, as an example, if two metrics differ only for the $g_{rr}$ component, they cannot be clearly distinguished by this method \cite{Defalco2021wh}.

The aforementioned investigation must be complemented with further analyses. As shown in Fig.~\ref{fig:Fig2}, the radial epicyclic frequency reveals some discrepancies in the close vicinity of the ISCO radius, up to $\sim7M$, and of course far from the gravitational source. Besides being straightforward, this procedure involves simultaneously $g_{tt}$ and $g_{rr}$, being therefore sensible in detecting departures in both metric components.

To conclude, we draw the BH shadow profile in  Fig.~\ref{fig:Fig3}, which simply reduces to a circle with radius $b_c$. Compared to the techniques adopted in the previous sections, however, this approach has the advantage to inquiry gravity in stronger gravitational field regimes. In the present study, the radii of the two shadows exhibit a relative gap of the order of $1\%$, which thus does not provide significant information. On the other hand, determining the ISCO position turns out to be extremely important, as it allows to place constraints on the metric tensor, with an associated error of the order of $5.5\%$. Furthermore, by increasing the inclination angle of the observer, the relativistic effects are magnified, thus permitting to viably disclose the BH metric properties. 

Therefore, the present analysis, based on three complementary astrophysical methods, is very solid for inquiring about the characteristics of the spacetime geometry around a static and spherically symmetric BH. From the observational point of view, we have seen that the current data on epicyclic frequencies are already sufficiently precise for detecting metric departures, whereas the required accuracy of EHT data (involving the image of a BH and the shadow profile) needs further sensitivity upgrades.

\section{Probing the solutions in presence of gravity and radiation}
\label{sec:PR effect}
The present section is devoted to the study of metric deviations occurring in presence of both the BH gravitational field and external forces generated by electromagnetic radiation processes. If the accreting matter is assumed to be made of uncharged test particles, the relativistic PR effect can be suitably exploited.

First, we shall introduce the gravitational setup to investigate the main features of the radiation stress-energy tensor. Then, we describe the test particle motion and its interaction with the radiation field. Based on these premises, we thus write the relativistic PR equations of motion. We eventually show that the trajectories driven by the relativistic PR effect can be used to distinguish the Schwarzschild metric from other similar geometries.

\subsection{Gravitational setup}
\label{sec:geometry}
We consider a static BH, whose external spacetime geometry is described by the metric \eqref{eq:metric2}. Setting $\theta = \pi/2$, we consider the orthonormal frame for both a SO at infinity $(\boldsymbol{\partial_\nu})^\mu:=\delta^\mu_\nu$ and for a SO located around the BH.  In the latter case, the future-pointing unit vector orthogonal to the spatial hypersurfaces is given by $
\boldsymbol{n}=\frac{1}{N}\boldsymbol{\partial_t}\,$, where $N=\sqrt{-g_{tt}}$ is the lapse function. A possible set of tetrad fields adapted to the SO frame are 
\begin{equation} \label{eq:zamoframes}
\begin{aligned}
&\boldsymbol{e_{\hat t}}=\boldsymbol{n},\qquad
\boldsymbol{e_{\hat r}}=\frac1{\sqrt{g_{rr}}}\boldsymbol{\partial_r},\qquad \boldsymbol{e_{\hat \varphi}}=\frac{1}{r}\boldsymbol{\partial_\varphi}.
\end{aligned}
\end{equation}
{Hereafter, all quantities (such as, $v^\alpha$ and $T^{\alpha\beta}$)} associated to the SO frame will be labeled by a hat, while all the scalar quantities measured in the SO frame {(such as} $f$) will be followed by $(n)$ (\emph{e.g.}, $f(n)$). In the kinematic decomposition of the SO congruence, the nonzero quantities are the acceleration $\boldsymbol{a}(n)=\nabla_{\boldsymbol{n}} \boldsymbol{n}=a(n)^{\hat r}(n) \boldsymbol{e_{\hat r}}$ and the relative Lie curvature $\boldsymbol{k_{(\rm Lie)}}(n)=k_{(\rm Lie)}(n)^{\hat r}\boldsymbol{e_{\hat r}}$, whose general definitions are 
\begin{subequations}
\begin{align}
a(n)^{\hat r}(n)&=\frac{1}{\sqrt{g_{rr}}}\partial_r\ln(N),\\
k_{(\rm Lie)}(n)^{\hat r}&=-\frac{1}{\sqrt{g_{rr}}}\partial_r\ln(r)=-\frac{1}{r\sqrt{g_{rr}}}.
\end{align}
\end{subequations}
The above expressions are reported in Table \ref{tab:Table2}, together with some other useful quantities.
\renewcommand{\arraystretch}{3}
\begin{table}[ht!]
    \caption{Physical quantities related to the BH solution ($\epsilon=-0.01$) compared with Schwarzschild metric's parameters. The BH mass is set to unity.}
    \centering
    \scalebox{1}{
    \begin{tabular}{|c|c|c|}
    \hline
    {\rm\bf Parameter} & {\rm\bf Our BH solution} & {\rm\bf Schwarschild}  \\
    \hline
    $N$ & $\displaystyle\left(\frac{1}{r^{0.02}}-\frac{2.02}{r^{1.03}}\right)^{1/2}$ & $\displaystyle\left(1-\frac{2}{r}\right)^{1/2}$ \\
    $a(n)^{\hat r}$ & $\displaystyle\frac{0.50 \left(\frac{2.08}{r^{2.03}}-\frac{0.02}{r^{1.02}}\right)}{\sqrt{\frac{1}{1-\frac{2.02}{r^{1.01}}}} \left(\frac{1}{r^{0.02}}-\frac{2.02}{r^{1.03}}\right)}$ & $\displaystyle\frac{1}{r^2}\left(1-\frac{2}{r}\right)^{-1/2}$\\    
    $\displaystyle k_{(\rm Lie)}(n)^{\hat r}$ & $\displaystyle-\frac{0.99}{r}\sqrt{{1-\frac{2.02}{r^{1.01}}}}$ & $\displaystyle-\frac{1}{r}\left(1-\frac{2}{r}\right)^{1/2}$ \\   $E(n)$ & $\displaystyle\frac{E}{\sqrt{\frac{1}{r^{0.02}}-\frac{2.02}{r^{1.03}}}}$ & $\displaystyle E\left(1-\frac{2}{r}\right)^{-1/2}$ \\
    $\cos\beta$ & $\displaystyle\frac{b}{r} \sqrt{\frac{1}{r^{0.02}}-\frac{2.02}{r^{1.03}}}$ & $\displaystyle\frac{b}{r}\left(1-\frac{2}{r}\right)^{1/2}$ \\ 
    $\mathcal{I}$ & $\displaystyle\frac{\mathcal{I}_0^2\left(\frac{1}{r^{0.02}}-\frac{2.02}{r^{1.03}}\right)^{-1/2}}{r \sqrt{r^2-b^2 \left(\frac{1}{r^{0.02}}-\frac{2.02}{r^{1.03}}\right)}}$ & $\displaystyle\frac{\mathcal{I}_0^2}{\sqrt{r^2-b^2\left(1-\frac{2}{r}\right)}}$\\
    $b(\RS,\Omega_\star)$ & $\displaystyle\left(\frac{\RS^2}{\frac{1}{\RS^{0.02}}-\frac{2.02}{\RS^{1.03}}}\right)\Omega_\star$ & $\displaystyle \left(\frac{\RS^2}{1-2/\RS}\right)\Omega_\star$ \\
    $\Omega_{\rm max}(\RS)$ & $\displaystyle \frac{1}{\RS}\sqrt{\frac{1}{\RS^{0.02}}-\frac{2.02}{\RS^{1.03}}}$ & $\displaystyle \frac{\sqrt{1-2/\RS}}{\RS}$ \\
    \hline    
    \end{tabular}}
    \label{tab:Table2}
\end{table}

\subsection{Stress-energy tensor for radiation}
\label{sec:photons}
The radiation field is treated as a coherent flux of photons travelling along null geodesics on the background spacetime \eqref{eq:metric2} and continuously hitting the test particle. The related stress-energy tensor is
\begin{equation}\label{STE}
T^{\mu\nu}=\mathcal{I}^2 k^\mu k^\nu\,,\qquad k^\mu k_\mu=0,\qquad k^\mu \nabla_\mu k^\nu=0,
\end{equation}
where $\mathcal{I}$ is  the radiation field intensity and $\boldsymbol{k}$ is the photon four-momentum field. By decomposing $\boldsymbol{k}$ with respect to the SO frame, we obtain
\begin{subequations} \label{photon}
\begin{align}
\boldsymbol{k}&:=E(n)[\boldsymbol{n}+\hat{\boldsymbol{\nu}}(k,n)],\\
\hat{\boldsymbol{\nu}}(k,n)&:=\sin\beta\ \boldsymbol{e_{\hat r}}+\cos\beta\ \boldsymbol{e_{\hat\varphi}},
\end{align}
\end{subequations}
with $\hat{\boldsymbol{\nu}}(k,n)$ being the photon spatial unit relative velocity with respect to the SO, and $\beta$ is the angle measured in the SO frame in the azimuthal direction. The radiation field is governed by the impact parameter $b$, associated to the emission angle $\beta$. The photon energy $E(n)$ and the photon angular momentum along the $z$-axis (orthogonal to the equatorial plane) $L_{\hat z}(n)$, can be expressed in the rest-frame of a distant observer as \cite{Bini2009,Bini2011} 
\begin{subequations} 
\begin{align}
E(n)&:=\frac{E}{N}, \label{eq:energySO}\\
L_{\hat z}(n)&:= E(n)\cos\beta=\frac{L_z}{r},\label{eq:momentumSO}
\end{align}
\end{subequations} 
where $E:=-k_t>0$, $L_z:=k_\varphi$, and $b:=L_z/E$. The photon impact parameter $b$ can be associated to its relative angle $\beta$ in the SO frame as follows (cf. Eq.~\eqref{eq:momentumSO}):
\begin{equation} \label{ANG1}
\cos\beta=\frac{b N}{r}.
\end{equation}
For $b=0$ (respectively, $b\neq0$) we model a radial (respectively, general) radiation field. The photon impact parameter $b$ can be written in terms of physical quantities, such as  the radius $\RS$ and the angular velocity $\Omega_\star$ of the emitting surface, namely \cite{Bakala2019}
\begin{equation}\label{eq:explicit_impact_parameter}
b=\frac{\RS^2 }{-g_{tt}\mathrm{(\RS)}}\Omega_{\star},
\end{equation} 
with the maximum angular velocity given by
\begin{equation}
\Omega_{\rm max}(\RS):=\frac{\sqrt{-g_{tt}(\RS)}}{\RS}.
\end{equation}
Notice that the condition $\Omega>\Omega_{\rm max}$ is not physically viable, as it leads to superluminal velocities.

From the conservation equation $\nabla_\mu(\mathcal{I}^2\sqrt{-g})=0$, it is possible to determine the intensity parameter as \cite{Bini2009,Bini2011} 
\begin{equation}
\mathcal{I}^2=\frac{\mathcal{I}_0^2}{r\sqrt{r^2-b^2N^2}}.    
\end{equation}

\subsection{Test particle motion}
\label{sec:test_particle}
Let us now consider a test particle moving in the equatorial plane with timelike four-velocity $\boldsymbol{U}$ and spatial velocity $\boldsymbol{\nu}(U,n)$ with respect to the SO frame given by, respectively,
\begin{subequations}
\begin{align}
\boldsymbol{U}&:=\gamma(U,n)[\boldsymbol{n}+\boldsymbol{\nu}(U,n)], \label{eq:testp}\\
\boldsymbol{\nu}&:=\nu(\sin\alpha\ \boldsymbol{e_{\hat r}}+\cos\alpha\ \boldsymbol{e_{\hat\varphi}}).
\end{align}
\end{subequations} 
In the above equations, $\gamma\equiv\gamma(U,n)=1/\sqrt{1-||\boldsymbol{\nu}(U,n)||^2}$ is the Lorentz factor, $\nu=||\boldsymbol{\nu}(U,n)||$ is the magnitude of the test particle spatial velocity $\boldsymbol{\nu}(U,n)$, and $\alpha$ is the azimuthal angle of the vector $\boldsymbol{\nu}(U,n)$ measured in the SO frame clockwise from the positive direction in the $\hat{r}-\hat{\varphi}$ tangent plane (see Fig.~\ref{fig:Fig_astro}). The explicit expression for the test particle velocity components with respect to the SO are
\begin{equation} 
\begin{aligned}\label{four_velocity}
&U^t=\frac{\gamma}{N},\qquad U^r=\frac{\gamma\nu^{\hat r}}{\sqrt{g_{rr}}},\qquad U^\varphi=\frac{\gamma\nu^{\hat\varphi}}{r},
\end{aligned}
\end{equation}
where $\tau$ is the proper time parameter along $\bold{U}$. 

By means of the \emph{observer-splitting formalism}, the test particle acceleration with respect to the SO frame, \emph{i.e.}, $\boldsymbol{a}(U)=\nabla_{\bold U} \bold{U}$, can be expressed as \cite{Bini2009,Bini2011} 
\begin{subequations}
\begin{align}
a(U)^{\hat t}=& \gamma^2\nu\sin\alpha\ [a(n)^{\hat r}+k_{\rm (Lie)}(n)^{\hat r}\,\nu^2\cos^2\alpha]\notag\\
&+\gamma^3 \nu\frac{\rm d \nu}{\rm d \tau},\label{eq:acc_t}\\   
a(U)^{\hat r}=& \gamma^2[a(n)^{\hat r}+k_{\rm (Lie)}(n)^{\hat r}\,\nu^2\cos^2\alpha]\notag\\
&+\gamma\left(\gamma^2\sin\alpha\frac{\rm d\nu}{\rm d\tau}+ \nu\cos\alpha\frac{\rm d \alpha}{\rm d \tau}\right),\label{eq:acc_r}\\ 
a(U)^{\hat \varphi}=& -\gamma^2 \nu^2\cos \alpha\sin\alpha k_{\rm (Lie)}(n)^{\hat r}\notag\\ 
&+ \gamma\left(\gamma^2 \cos \alpha\frac{\rm d \nu}{\rm d \tau}-\nu\sin \alpha\frac{\rm d \alpha}{\rm d \tau}\right).\label{eq:acc_varphi}
\end{align}
\end{subequations}

\subsection{Radiation force field}
\label{sec:radiation_field}
We assume that the radiation-test particle interaction occurs through an elastic Thomson-like scattering, characterized by a constant momentum-transfer cross section $\sigma$, which is assumed to be independent of the direction and frequency of the radiation field. Therefore, the radiation force is \cite{Bini2009,Bini2011} 
\begin{equation} \label{radforce}
{\mathcal F}_{\rm (rad)}(U)^\alpha := -\sigma P(U)^\alpha{}_\beta \, T^{\beta}{}_\mu \, U^\mu \,,
\end{equation}
where $P(U)^\alpha{}_\beta=\delta^\alpha_\beta+U^\alpha U_\beta$ projects a given vector in the spatial hypersurface orthogonal to $\boldsymbol{U}$. Decomposing the photon four-momentum $\boldsymbol{k}$ with respect to the test particle four-velocity $\boldsymbol{U}$ and considering the SO frame, we obtain the relation
\begin{equation} \label{diff_obg}
\boldsymbol{k}= E(n)[\boldsymbol{n}+\hat{\boldsymbol{\nu}}(k,n)]=E(U)[\boldsymbol{U}+\hat{\boldsymbol{\mathcal V}}(k,U)],
\end{equation}
which, substituted into Eq.~(\ref{radforce}), leads to
\begin{equation} \label{Frad0}
{\mathcal F}_{\rm (rad)}(U)^\alpha=\sigma \, [\Phi E(U)]^2\, \hat {\mathcal V}(k,U)^\alpha\,.
\end{equation}
The equations of motion for test particles are $m \bold{a}(U) = \boldsymbol{{\mathcal F}_{\rm (rad)}}(U)$,
with $m$ being the test particle mass. By defining $\tilde \sigma=\sigma/m$, we obtain
\begin{equation}\label{geom}
\bold{a}(U)=\tilde \sigma \Phi^2 E(U)^2  \,\hat{\boldsymbol{\mathcal V}}(k,U).
\end{equation} 
The scalar product between Eq.~(\ref{diff_obg}) and $\boldsymbol{U}$ provides 
\begin{equation} \label{enepart}
E(U)=\gamma E(n)[1-\nu\sin\psi\cos(\alpha-\beta)].
\end{equation}
The components of $\hat{\boldsymbol{\mathcal{V}}}(k,U)=\hat{\mathcal{V}}^t\boldsymbol{n}+\hat{\mathcal{V}}^r\boldsymbol{e_{\hat r}}+\hat{\mathcal{V}}^\varphi \boldsymbol{e_{\hat\varphi}}$ are
\begin{subequations}
\begin{align}
\hat{\mathcal{V}}^{\hat t}&=\gamma\nu\left[\frac{\cos(\alpha-\beta)-\nu}{1-\nu\cos(\alpha-\beta)}\right],\label{eq:rad_t}\\
\hat{\mathcal{V}}^{\hat r}&=\frac{\sin\beta}{\gamma [1-\nu\cos(\alpha-\beta)]}-\gamma\nu\sin\alpha,\label{eq:rad_r}\\
\hat{\mathcal{V}}^{\hat\varphi}&=\frac{\cos\beta}{\gamma [1-\nu\cos(\alpha-\beta)]}-\gamma\nu\cos\alpha.\label{eq:rad_varphi}
\end{align}
\end{subequations}

\subsection{Equations of motion}
\label{sec:EoM}
In light of the results of  previous sections, it is possible to write the relativistic PR equations of motion for a test particle moving in the equatorial plane as a set of coupled first-order differential equations: 
\begin{subequations}
\begin{align}
\frac{\dd\nu}{\dd\tau}&= -\frac{\sin\alpha}{\gamma}a(n)^{\hat r}+\frac{\tilde{\sigma}[\mathcal{I} E(U)]^2}{\gamma^3\nu}\hat{\mathcal{V}}^{\hat t},\label{eq:EoM1}\\
\frac{\dd\alpha}{\dd\tau}&=-\frac{\gamma\cos\alpha}{\nu}\left[a(n)^{\hat r}+k_{\rm (Lie)}(n)^{\hat r}\,\nu^2\right]\nonumber\\
&\ \quad+\frac{\tilde{\sigma}[\Phi E(U)]^2\cos\alpha}{\gamma\nu}\left[\hat{\mathcal{V}}^{\hat r}-\hat{\mathcal{V}}^{\hat \varphi}\tan\alpha\right],\label{eq:EoM2}\\
\frac{\dd r}{\dd\tau}&=\frac{\gamma\nu\sin\alpha}{\sqrt{g_{rr}}}, \label{eq:EoM3}\\
\frac{\dd\varphi}{\dd\tau}&=\frac{\gamma\nu\cos\alpha}{\sqrt{g_{\varphi\varphi}}}.\label{eq:EoM4}
\end{align}
\end{subequations}
The normalized luminosity of the radiation field is given by $A/M=\tilde{\sigma}\mathcal{I}_0^2E^2= L_\infty/L_{\rm Edd}$ and lies within the range $[0,1]$. Here, $L_\infty$ and $L_{\rm Edd}:=4\pi Mm/\sigma$ are the luminosity evaluated at infinity and the Eddington luminosity, respectively. Using Eq.~(\ref{enepart}), the term $\tilde{\sigma}[\mathcal{I} E(U)]^2$ can be recast as 
\begin{equation} \label{eq: sigma_tilde}
\tilde{\sigma}[\mathcal{I} E(U)]^2=\frac{ A\,\gamma^2[1-\nu\cos(\alpha-\beta)]^2}{N^2r^2|\sin\beta|}.
\end{equation}

\subsection{Critical surfaces}
\label{sec:critical_hypersurfaces}
The set of Eqs.~(\ref{eq:EoM1})--(\ref{eq:EoM4}) admits a region on which gravitational attraction, radiation pressure, and PR drag force balance. In the two-dimensional case, this region gives rise to a circle dubbed \emph{critical surface}, only characterized by its radius $r_{\rm (crit)}$. 

Therefore, let us consider a test particle in radial equilibrium, moving in a purely circular motion (\emph{i.e.}, $\alpha=0$ and $\nu=\nu_{\rm (crit)}$). Then, setting $ \frac{\dd\nu}{\dd\tau}=0$, Eq.~(\ref{eq:EoM1}) becomes
\begin{equation}\label{eq:crit_hyper1}
\frac{A(1 - \nu\cos\beta)(\cos\beta-\nu)}{N^2r^2|\sin\beta|}=0,
\end{equation}
from which we obtain $\nu_{\rm (crit)}=\cos\beta$. Thus, the velocity of the orbiting test particle is equal to the azimuthal velocity of the radiation field photons. It is worth noticing that, if the photons are radially emitted (that is $b=0$), then $\nu_{\rm (crit)}=0$ and the test particle will stop on a point on the critical surface. On the other hand, for $b\neq0$, the test particle will move on the critical surface with constant velocity $\nu_{\rm (crit)}$, being continuously pushed in the orbiting direction by the \qm{non-radial} photons of the radiation field. If we set $\alpha=\pi$ in Eq.~\eqref{eq:EoM1}, we obtain the same velocity $\nu_{\rm (crit)}$, but with a negative sign. In the SO frame, the test particle's velocity is tangential to the critical surface (\emph{i.e.}, $ \frac{\dd\alpha}{\dd\tau}=0$), so that Eq.~(\ref{eq:EoM2}) becomes
\begin{equation}\label{eq:crit_hyper2}
\begin{aligned}
&a(n)^{\hat r}+k_{\rm (Lie)}(n)^{\rm r}\nu_{\rm (crit)}^2 =\frac{A\sin\beta}{N^2\gamma^3 |\sin\beta|}.
\end{aligned}
\end{equation}
Therefore, once the values of the parameters $\left\{A,R_\star,\Omega_\star\right\}$ are assigned, the size of the critical surface can be obtained by solving Eq.~\eqref{eq:crit_hyper2} in the $r$-coordinate. 

\subsection{Trajectories of test particles}
\label{sec:PR_method}
Let us now take into account the test particle trajectories with viable values for the underlying relativistic PR variables.
In order to reduce the parameter space, we fix initial conditions on the test particle motion
by means of physical considerations. By examining the outcomes, we can infer information on how to detect deviations from the Schwarzschild geometry. An important feature of the relativistic PR effect, which will be extensively used, relies on its sensitive dependence on the initial conditions \cite{Defalco2021chaos}.

\begin{figure*}[ht!]
    \centering
    \hbox{
    \includegraphics[scale=0.26]{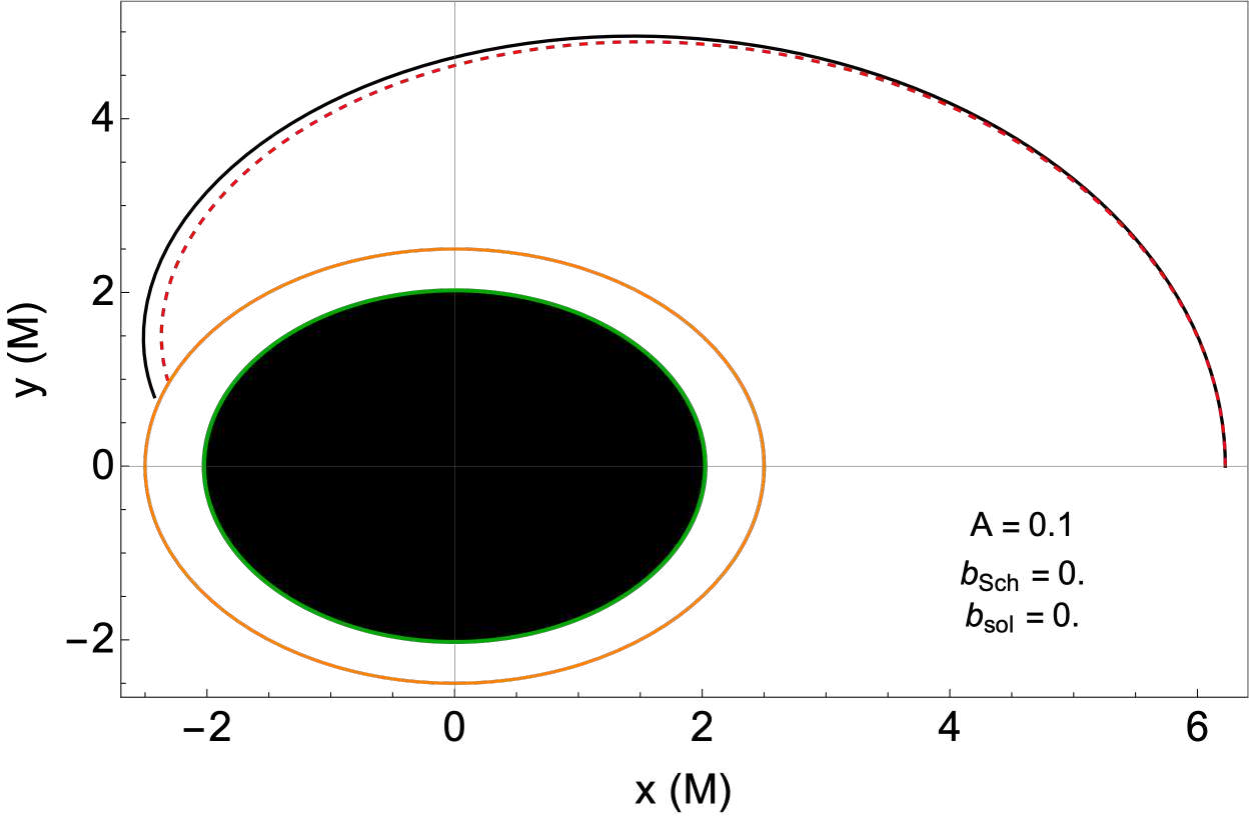}\hspace{0.3cm}
    \includegraphics[scale=0.26]{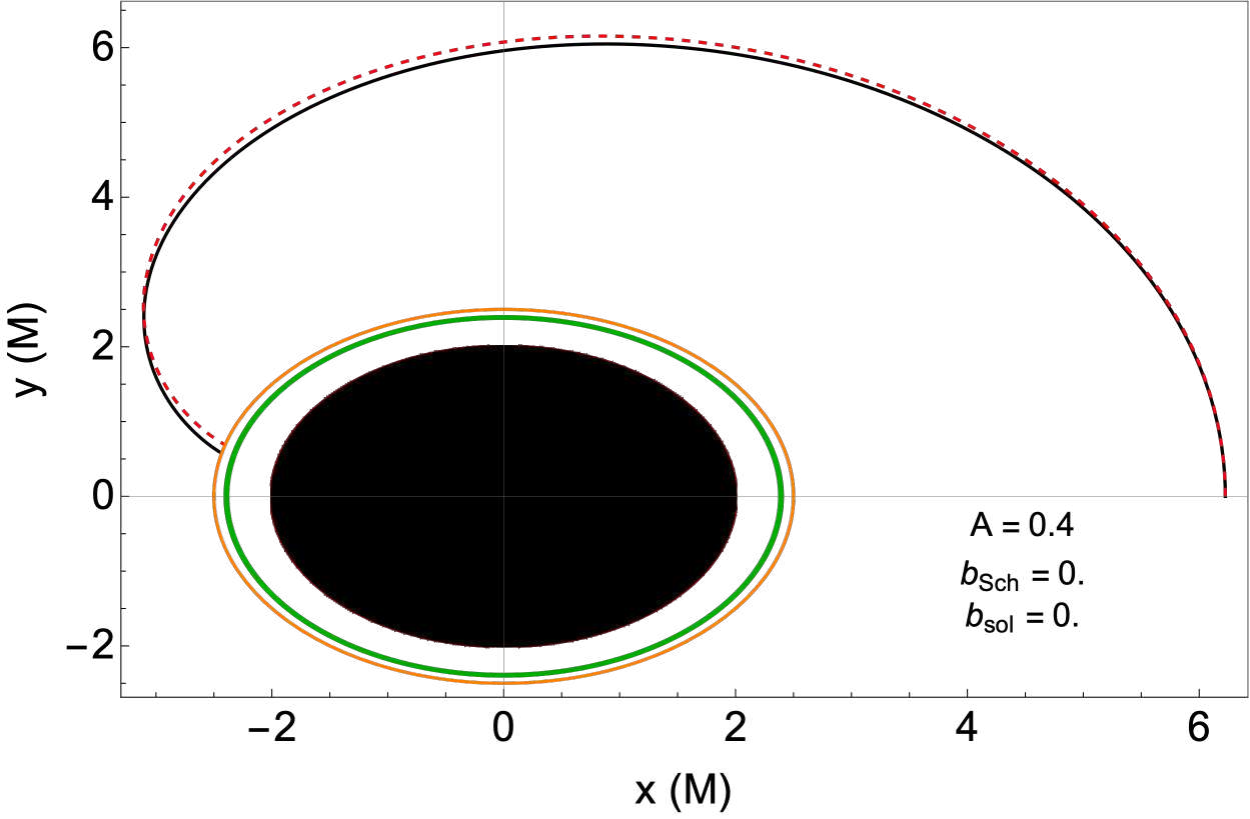}\hspace{0.3cm}
    \includegraphics[scale=0.26]{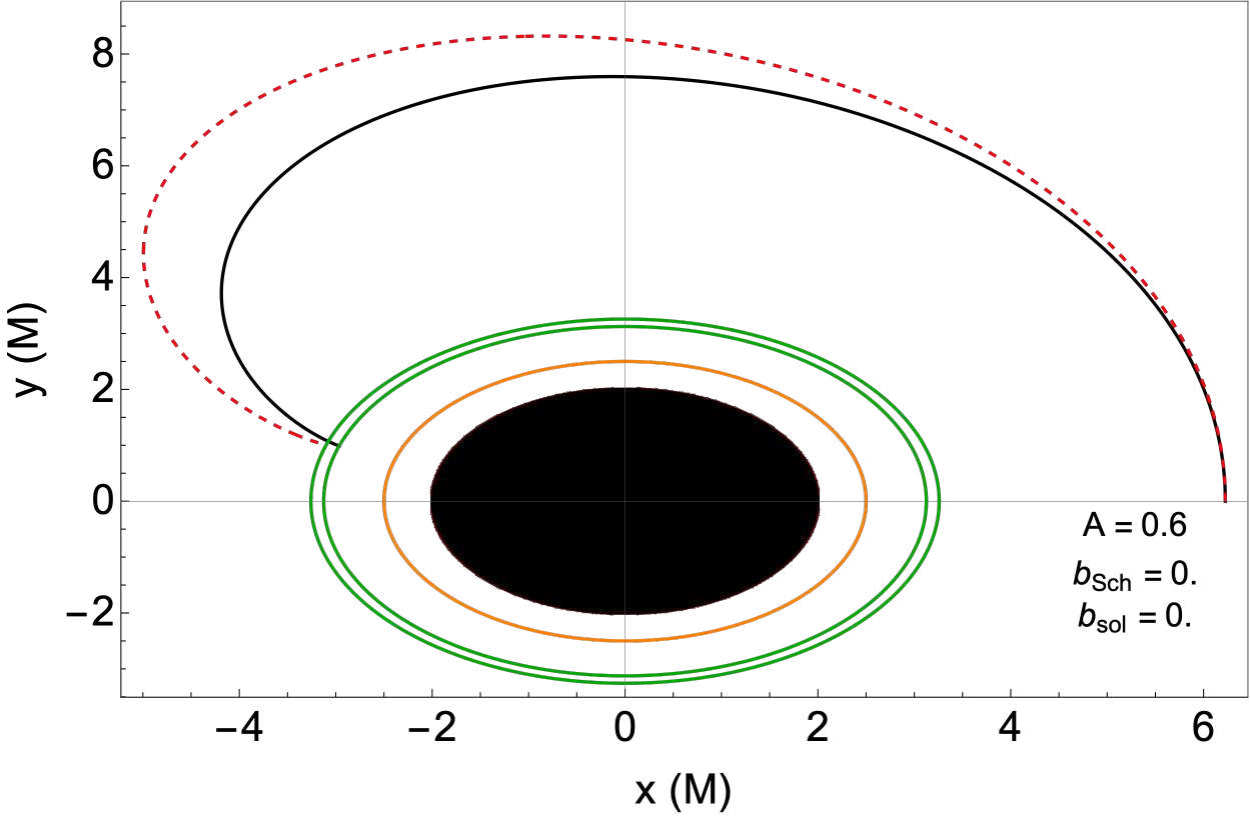}}\vspace{0.3cm}
    \hbox{
    \includegraphics[scale=0.26]{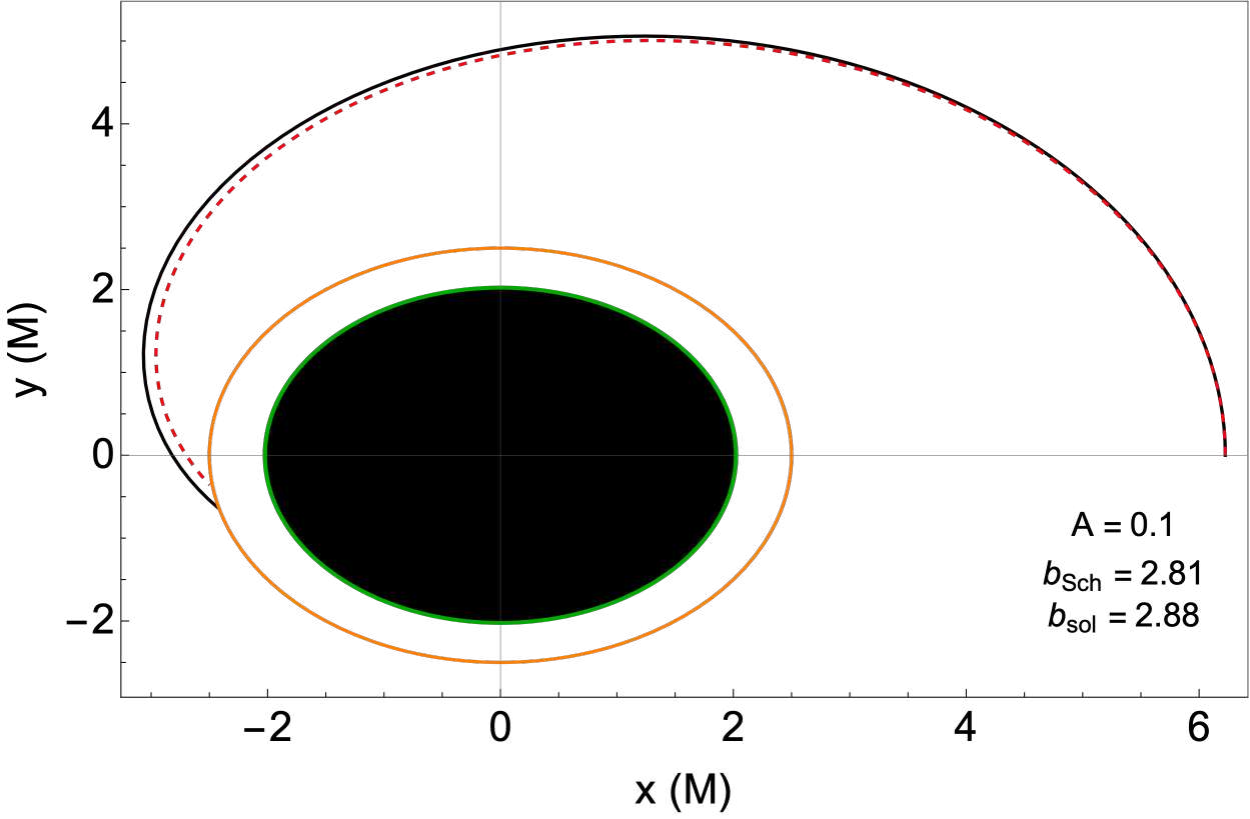}\hspace{0.3cm}
    \includegraphics[scale=0.26]{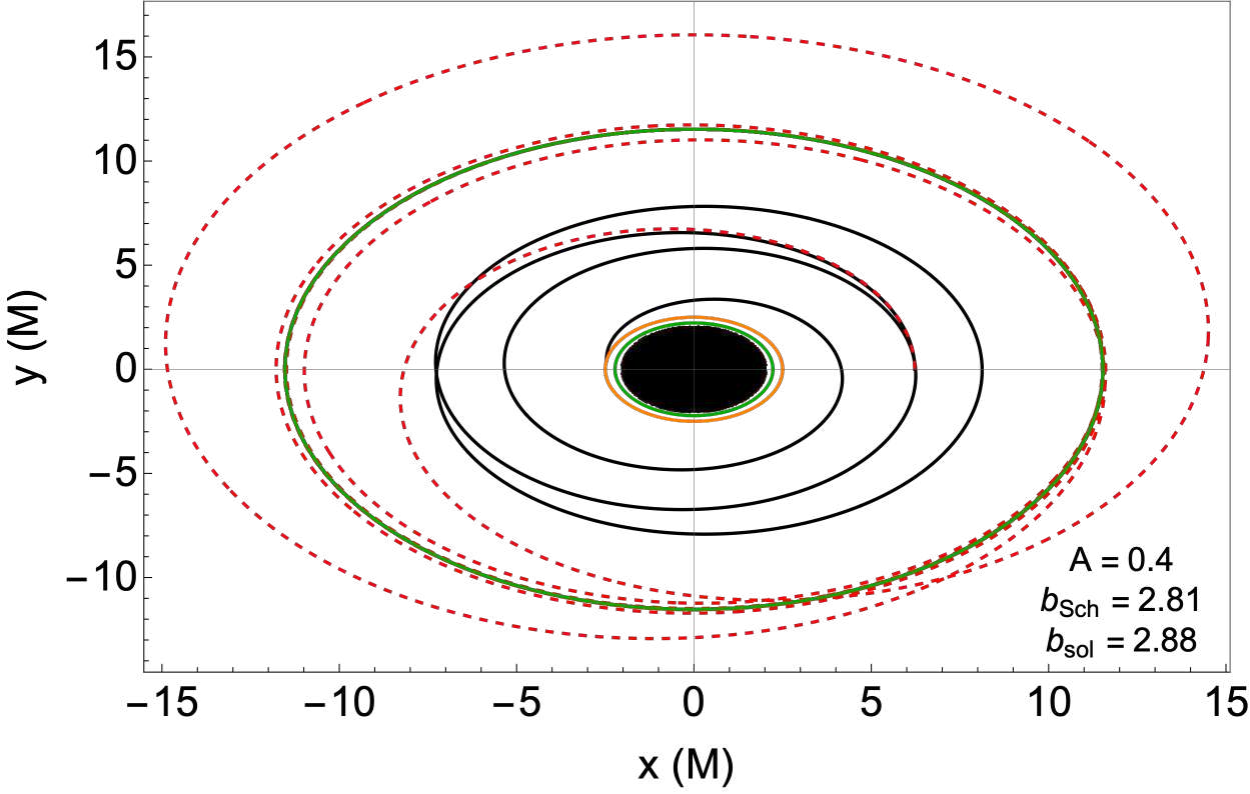}\hspace{0.3cm}
    \includegraphics[scale=0.26]{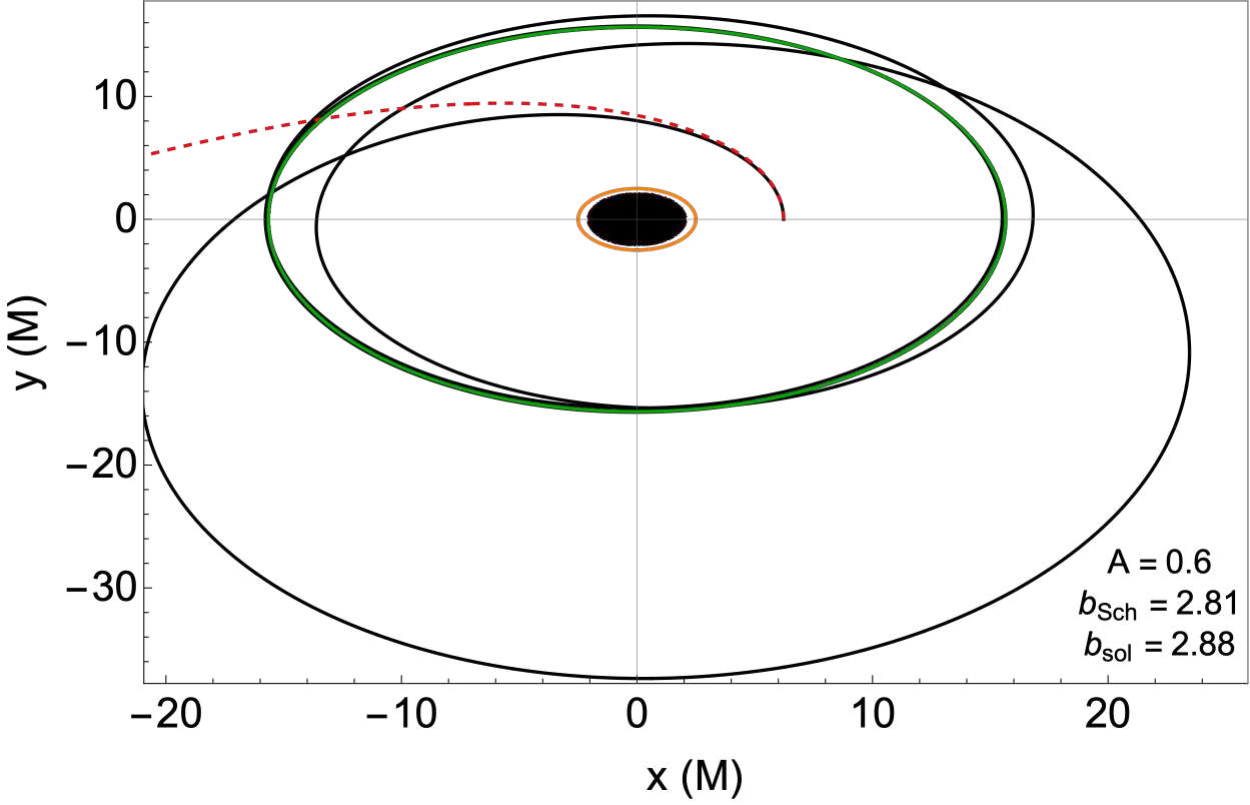}}\vspace{0.3cm}
     \hbox{
    \includegraphics[scale=0.26]{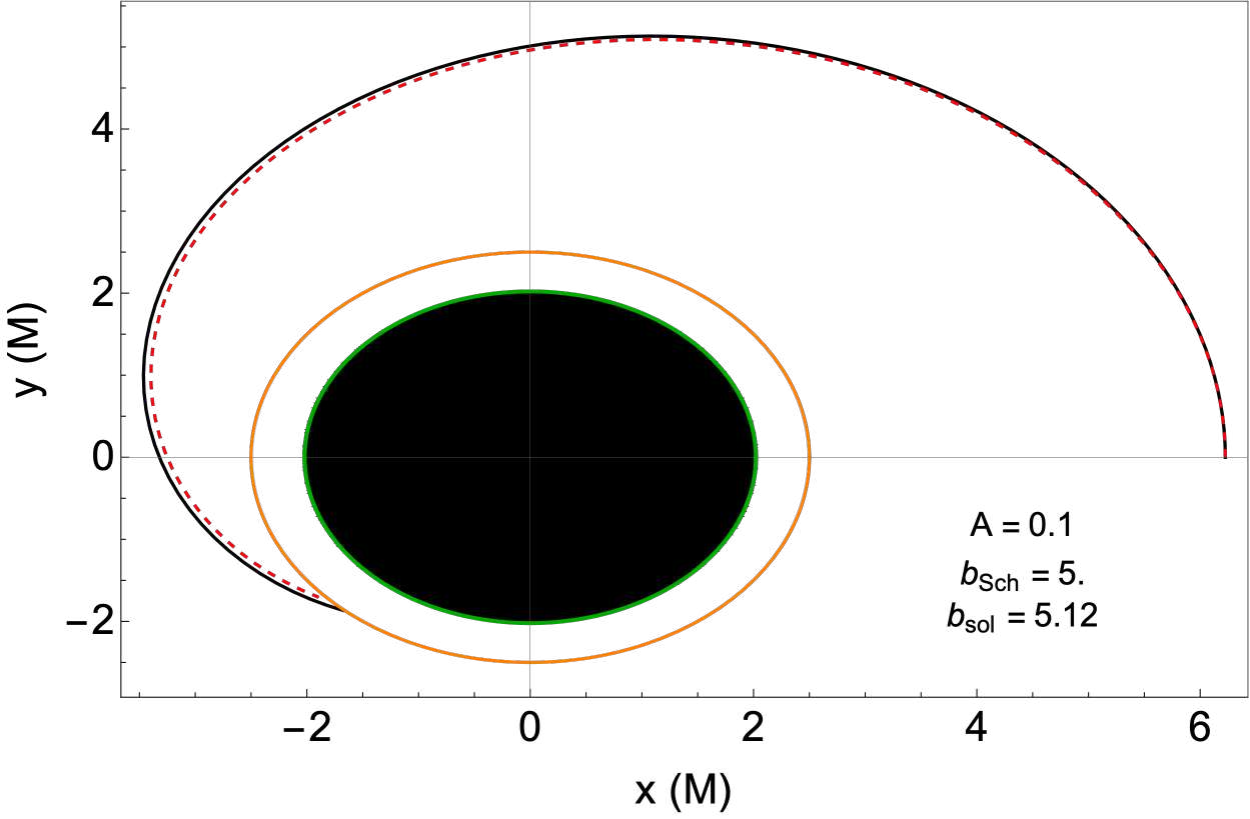}\hspace{0.3cm}
    \includegraphics[scale=0.26]{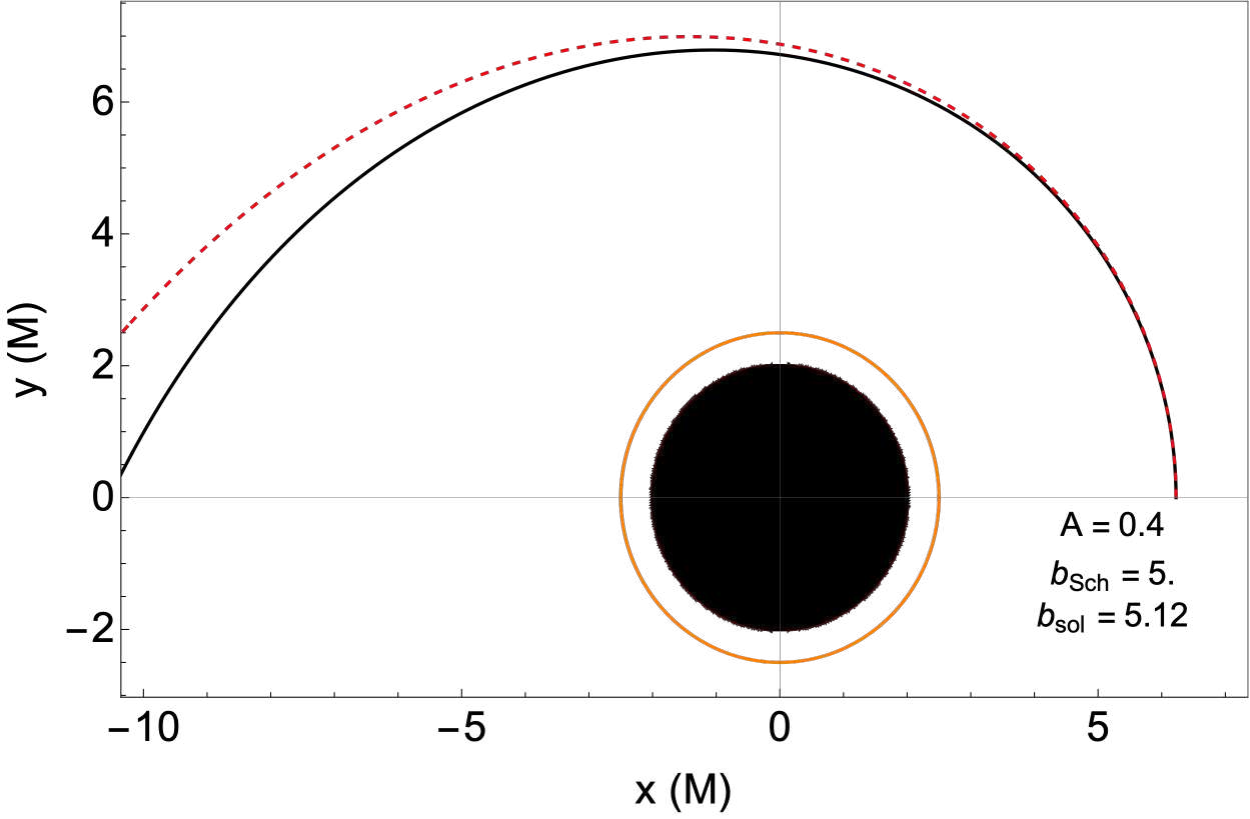}\hspace{0.3cm}
    \includegraphics[scale=0.26]{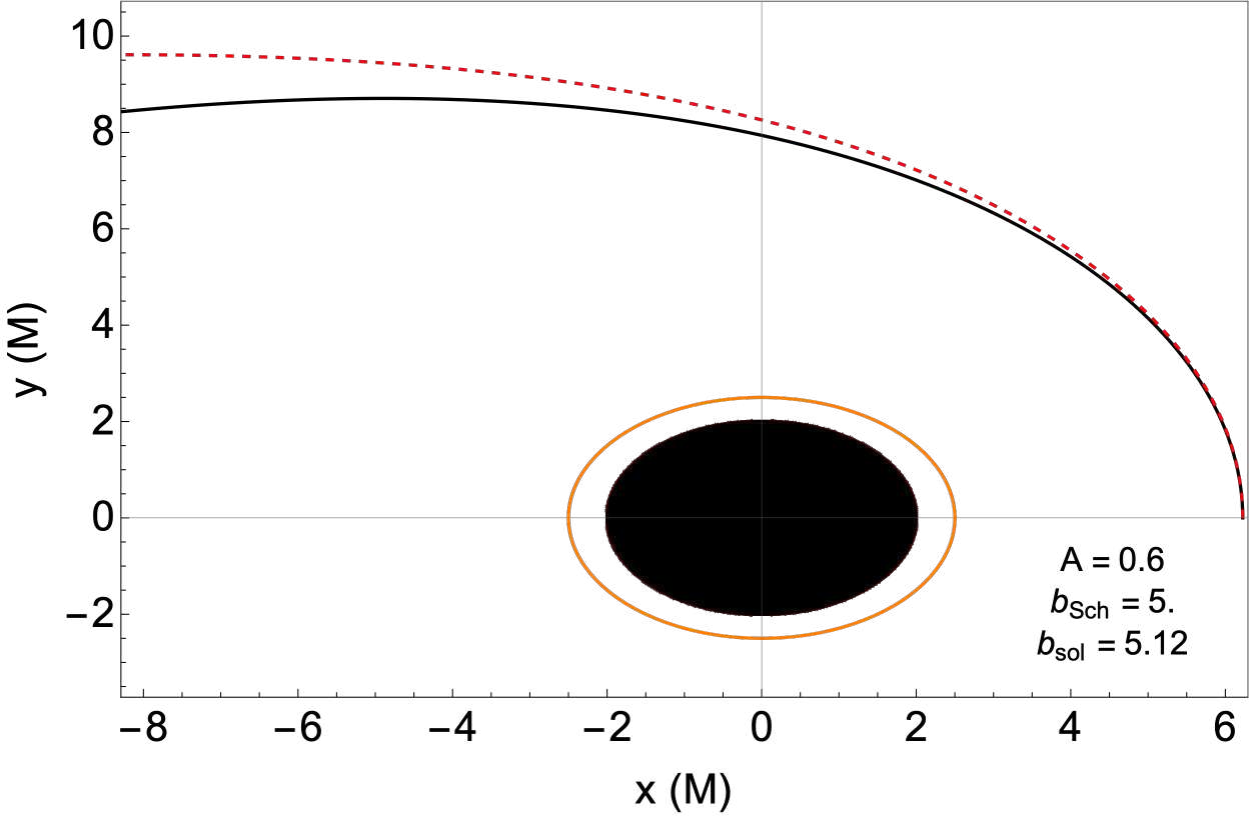}}
    \caption{Black trajectories refer to the Schwarzschild metric, whereas red dashed curves to our BH solution. The BH is placed at the center of the coordinate axes and its black boundary accounts for the Schwarzschild radius ($r_{\rm h}=2M$), while the red boundary for our BH solution ($r_{\rm h}=2.01M$). Due to the adopted scales, such boundaries are almost indistinguishable. The green and orange lines identify the critical region and the emitting surface, respectively. In each plot, we also report the values of the luminosity parameter $A$ and the photon impact parameters $b_{\rm Sch}$ and $b_{\rm sol}$. The initial conditions on the test particle are $(\nu_0,\alpha_0,r_0,\varphi_0)=(0.40,0,6.23M,0)$ in the Schwarzschild case and $(\nu_0,\alpha_0,r_0,\varphi_0)=(0.39,0,6.23M,0)$ for our BH solution.}
    \label{fig:FigPRa}
\end{figure*}

\begin{figure*}[ht!]
    \centering
    \hbox{
    \includegraphics[scale=0.26]{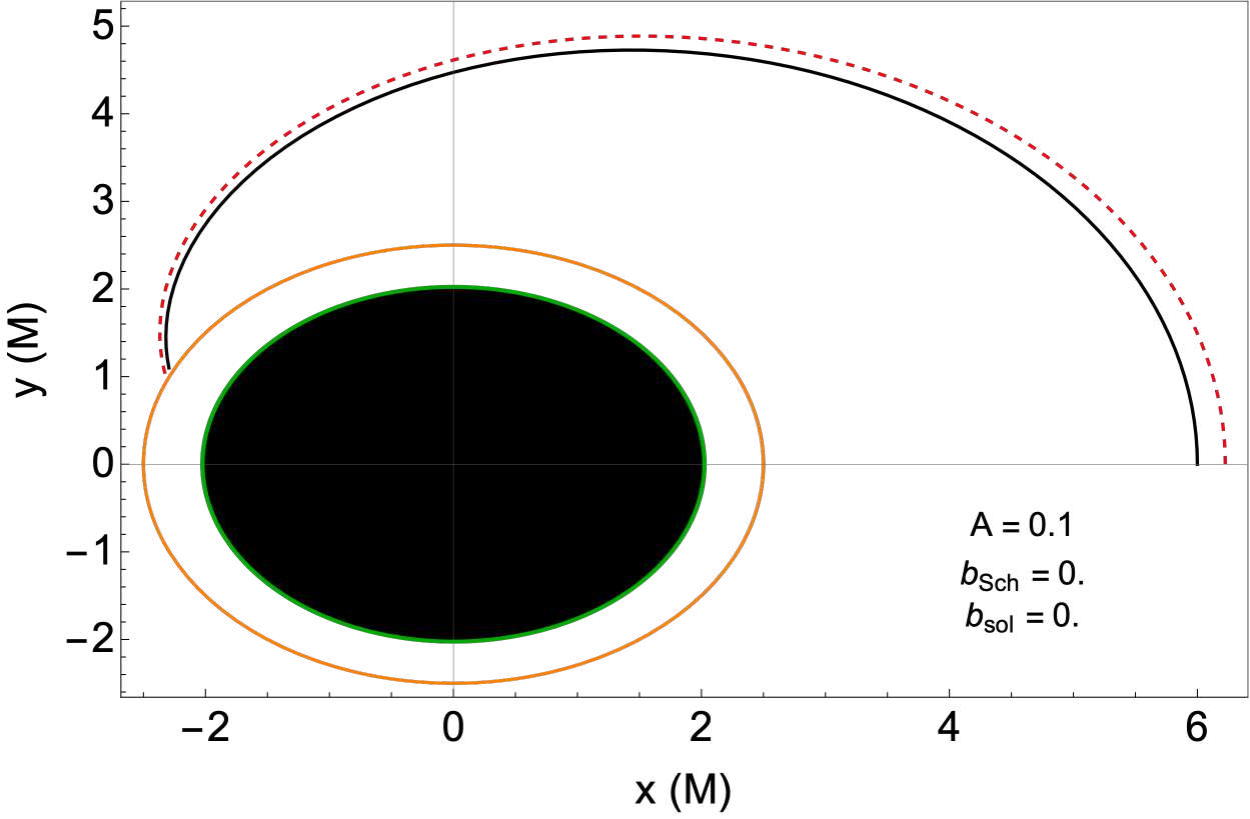}\hspace{0.3cm}
    \includegraphics[scale=0.26]{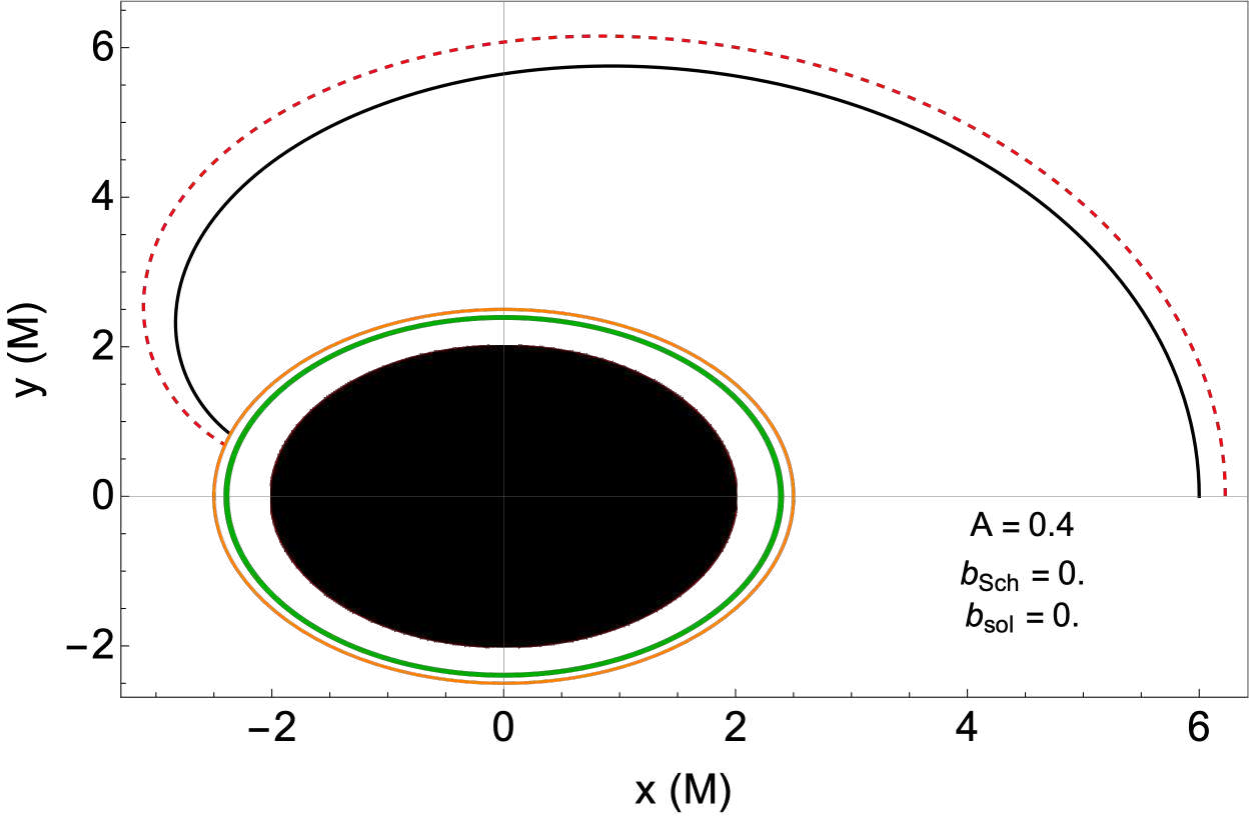}\hspace{0.3cm}
    \includegraphics[scale=0.26]{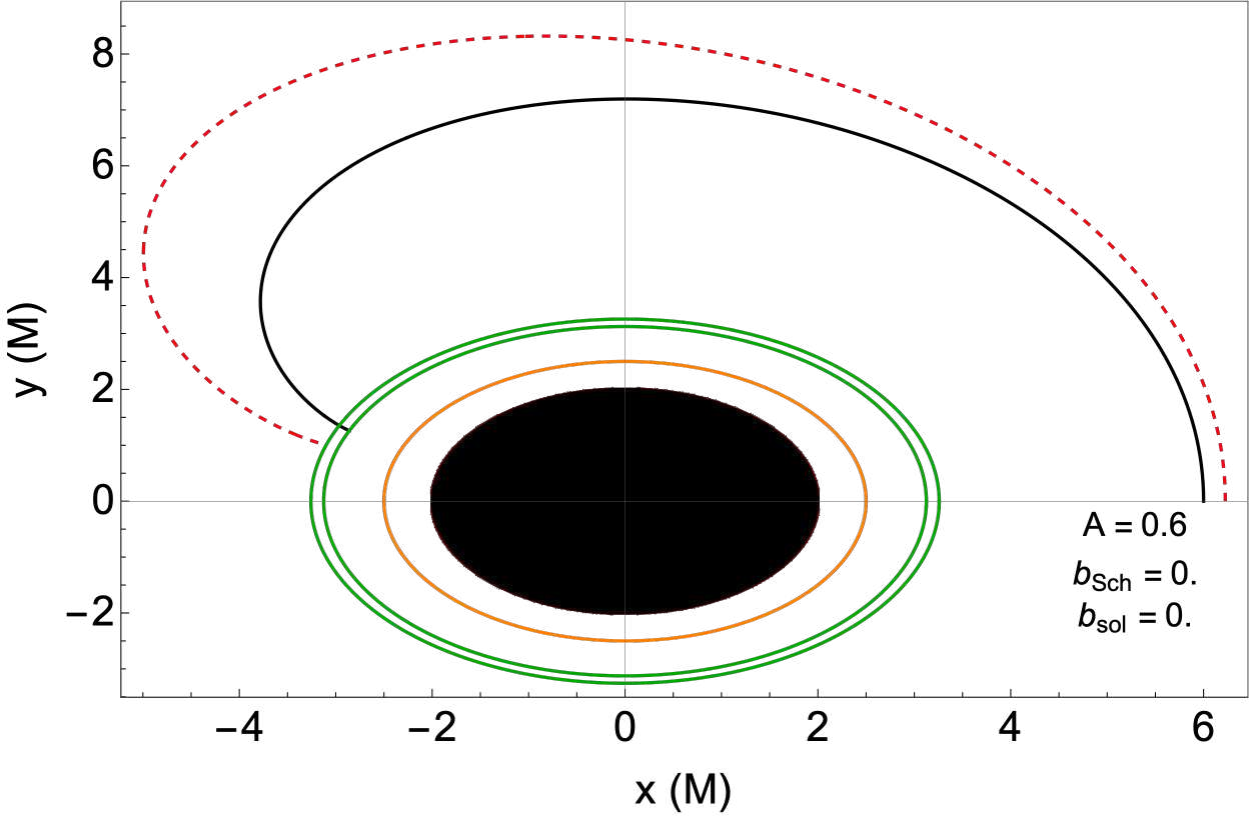}}\vspace{0.3cm}
    \hbox{
    \includegraphics[scale=0.26]{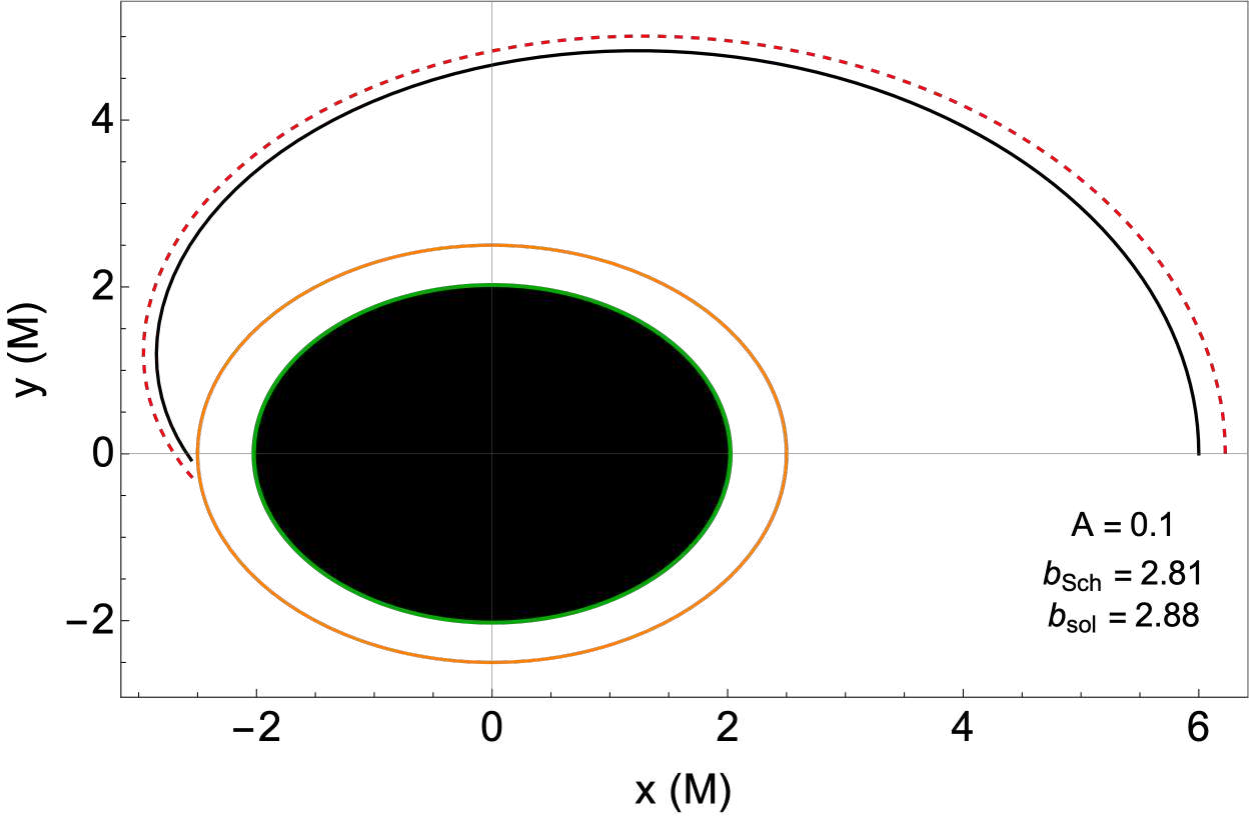}\hspace{0.3cm}
    \includegraphics[scale=0.26]{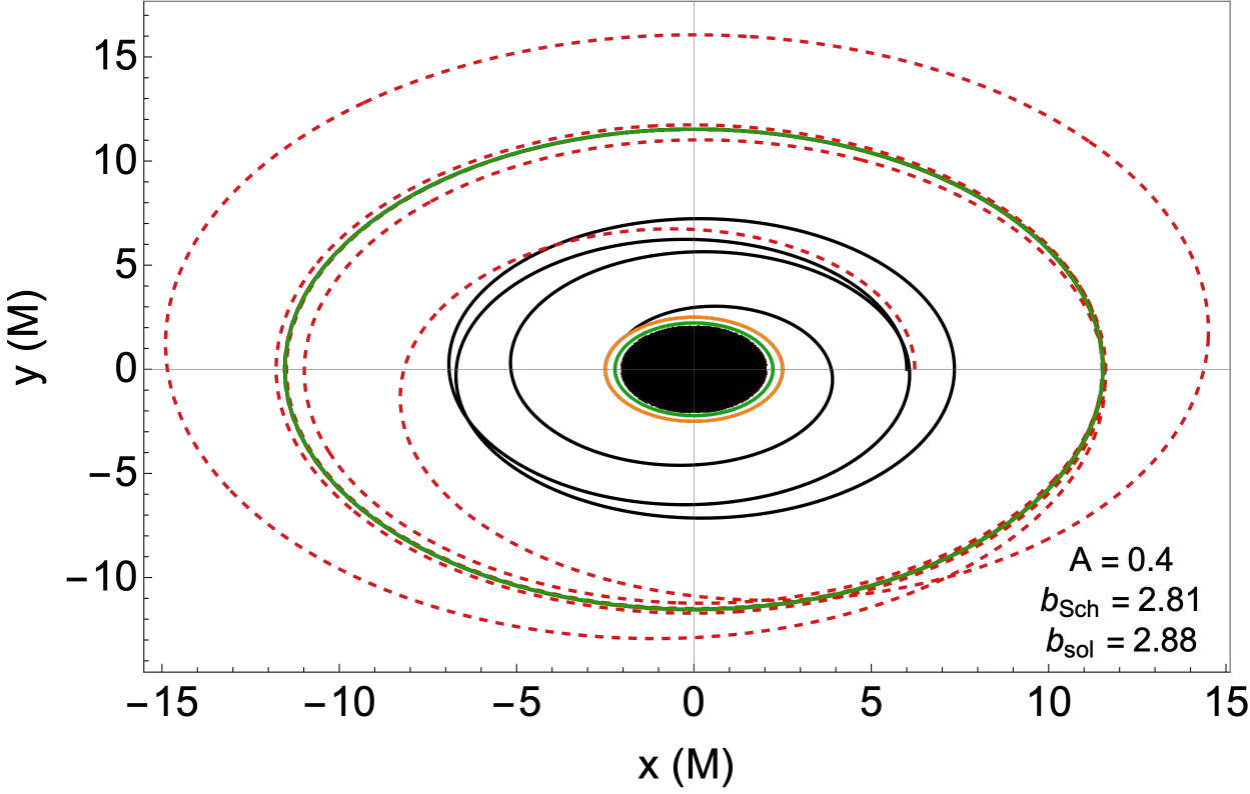}\hspace{0.3cm}
    \includegraphics[scale=0.26]{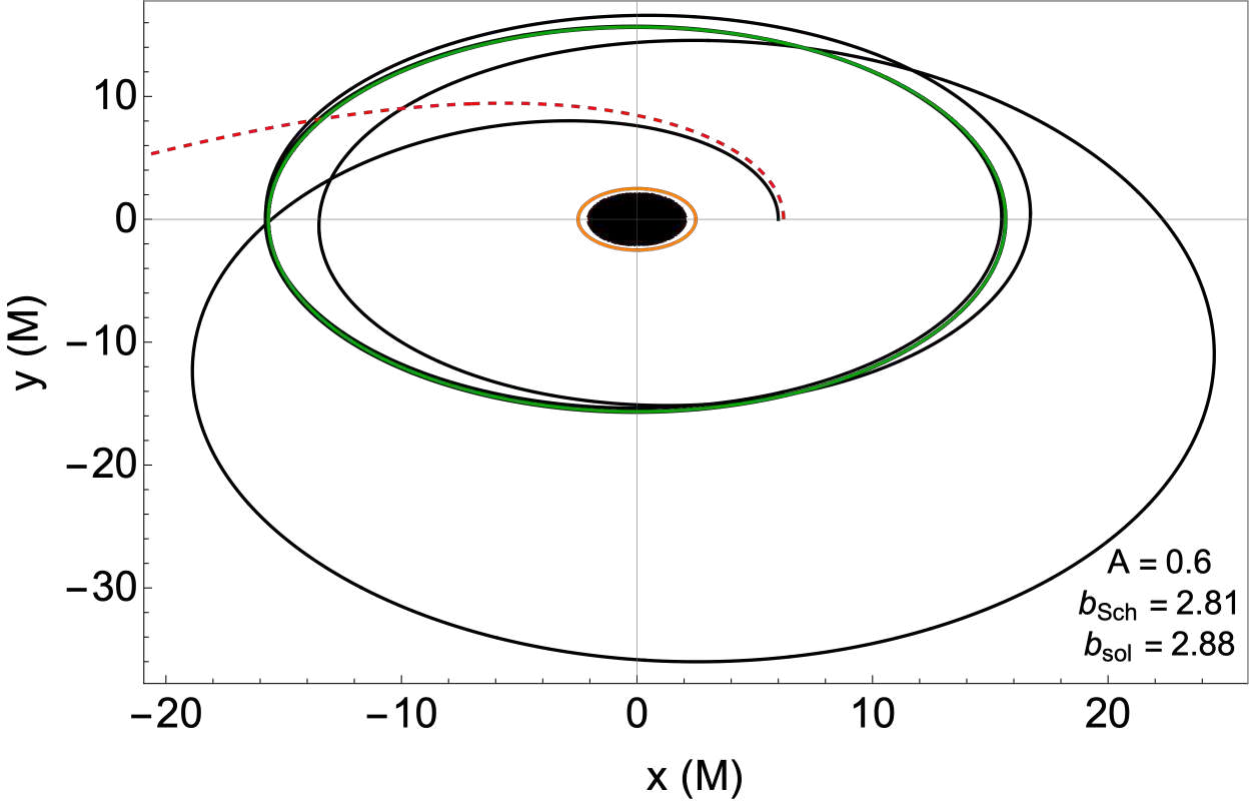}}\vspace{0.3cm}
    \hbox{
    \includegraphics[scale=0.26]{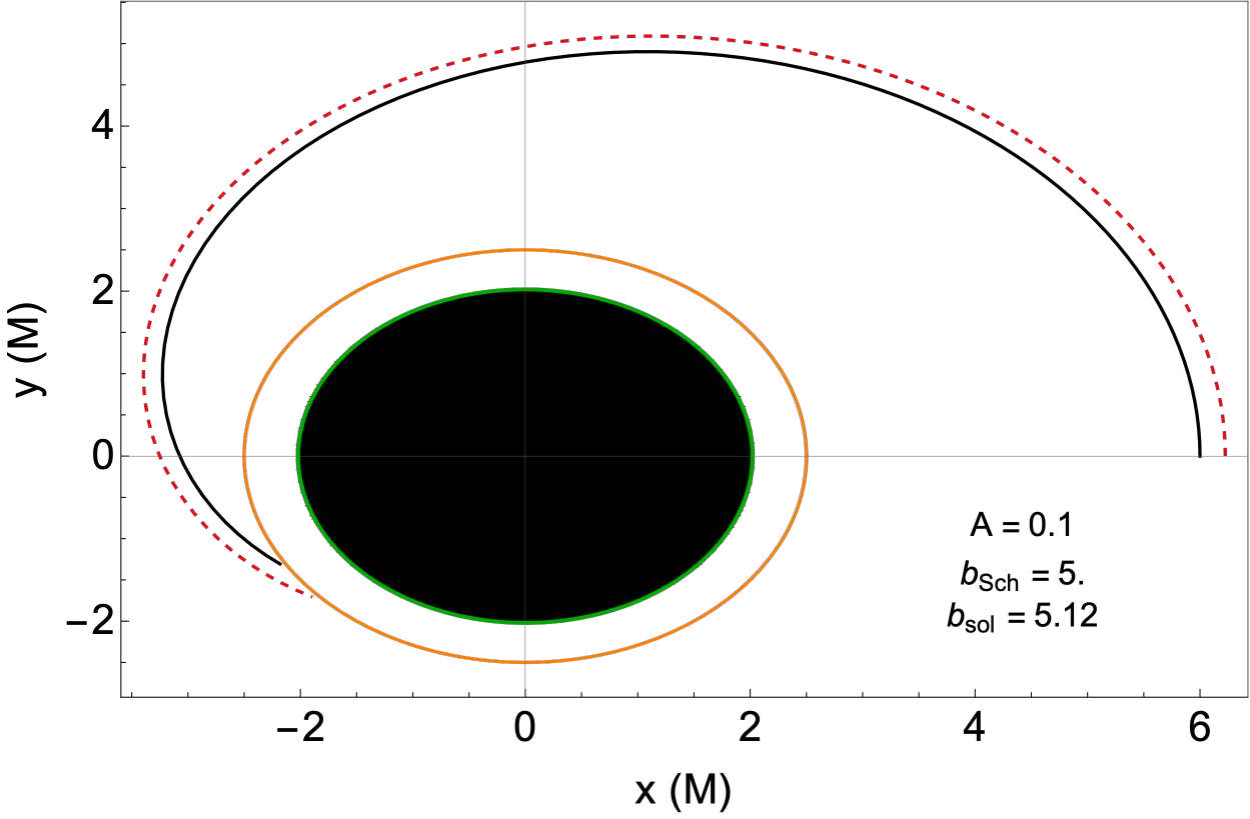}\hspace{0.3cm}
    \includegraphics[scale=0.26]{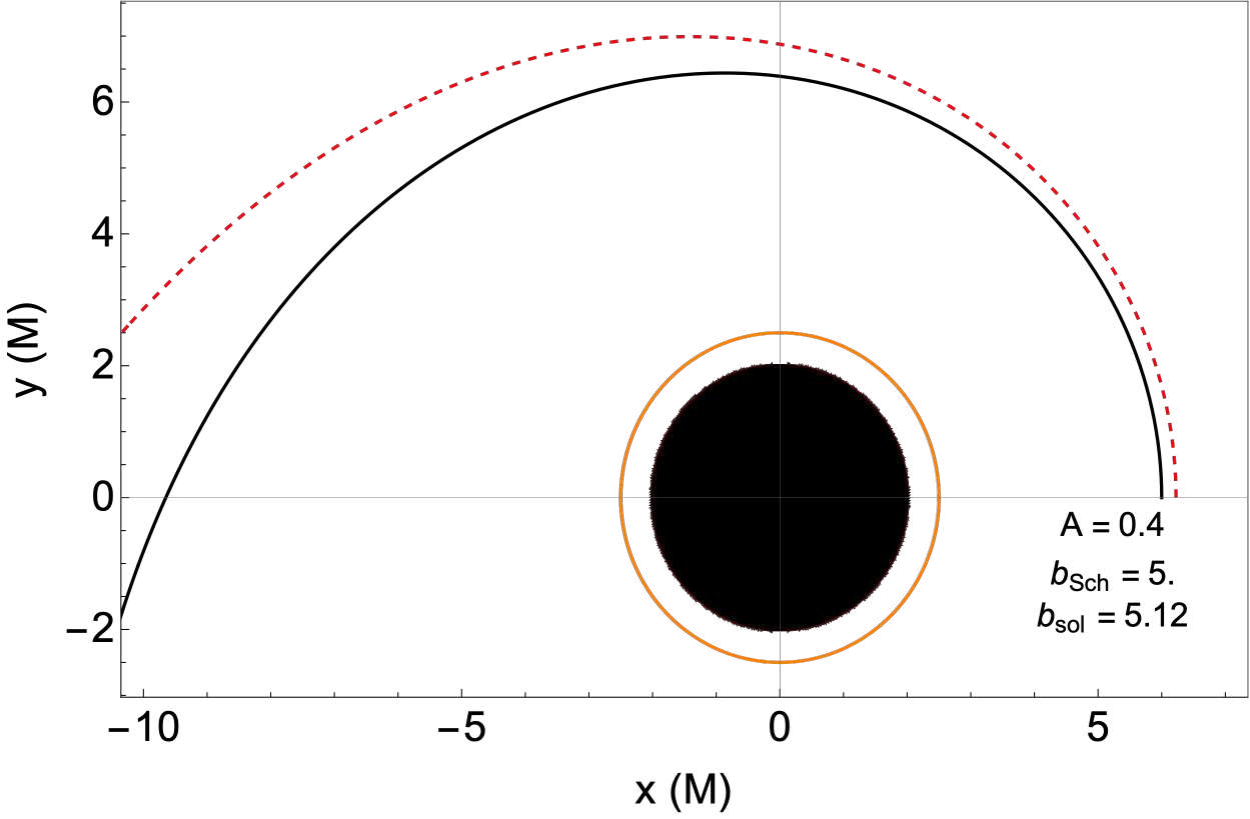}\hspace{0.3cm}
    \includegraphics[scale=0.26]{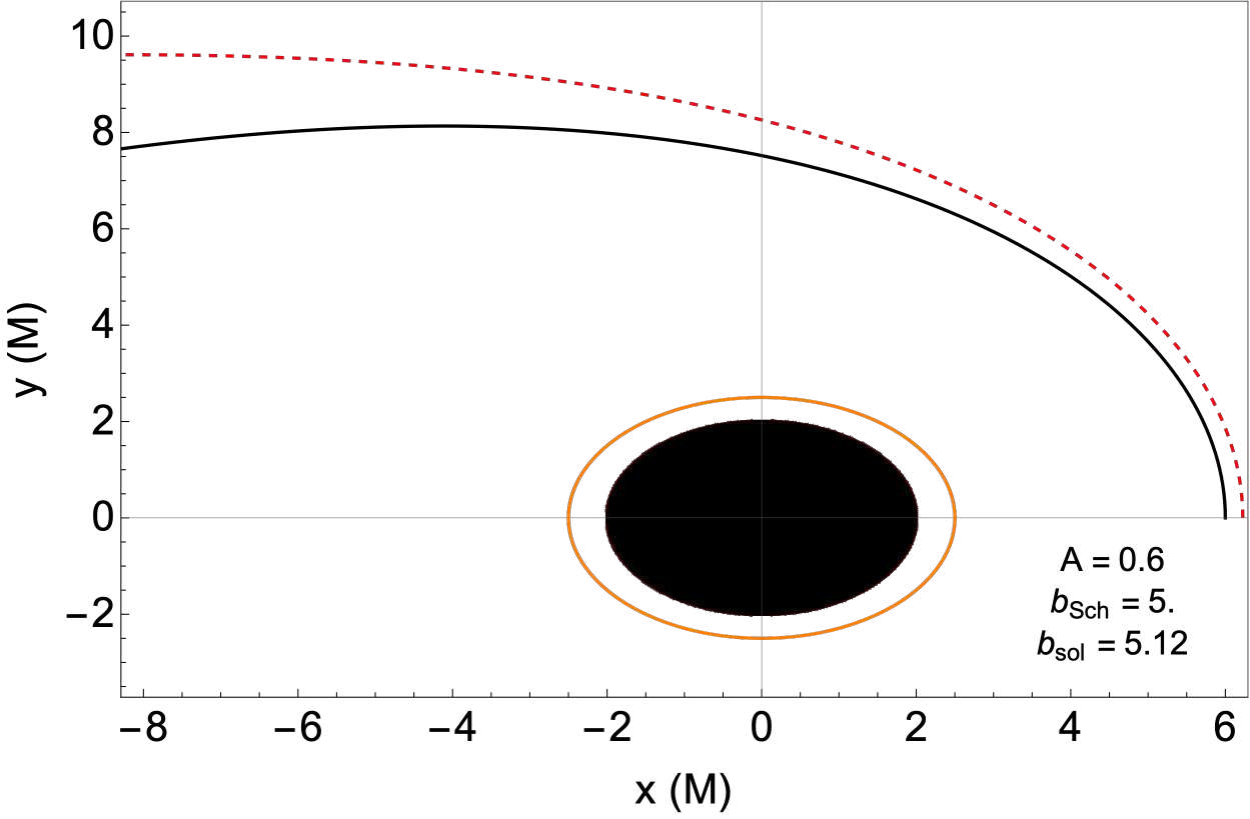}}
    \caption{The notation is the same as in Fig.~\ref{fig:FigPRa}. In this case, the test particle's initial conditions in the Schwarzschild metric are $(\nu_0,\alpha_0,r_0,\varphi_0)=(0.41,0,6M,0)$, whereas for our BH solution they are $(\nu_0,\alpha_0,r_0,\varphi_0)=(0.39,0,6.23M,0)$. }
    \label{fig:FigPR}
\end{figure*}

To develop the graphs in Fig.~\ref{fig:FigPRa}, test particles in all simulations are assumed to start from the same position, namely the ISCO radius of our BH solution and to move with Keplerian velocity. In other words, the initial conditions are fixed as $(\nu_0,\alpha_0,r_0,\varphi_0)=(\nu_K,0,r_{\rm ISCO},0)$, with $\nu_K=r_{\rm ISCO}\Omega_K(r_{\rm ISCO})$. It is worth noticing that, albeit the test particles start from the same radius, they have different velocities, because the expression of $\Omega_K$ is different in the two metrics. In Fig.~\ref{fig:FigPR}, the test particles start from the corresponding ISCO radius for both the metrics, but they always move with Keplerian velocity. Therefore, defining $\tilde{\nu}_K=\tilde{r}_{\rm ISCO}\Omega_K(\tilde{r}_{\rm ISCO})$, the initial conditions are fixed as $(\nu_0,\alpha_0,r_0,\varphi_0)=(\tilde{\nu}_K,0,\tilde{r}_{\rm ISCO},0)$, with $\tilde{r}_{\rm ISCO}=6M$ for Schwarzschild and $\tilde{r}_{\rm ISCO}=r_{\rm ISCO}$ for our BH solution. 

Regarding the radiation field variables, we choose three values for the luminosity parameter, \emph{i.e.}, $A=(0.1,0.4,0.6)$, corresponding to low, low-intermediate, and high-intermediate luminosities, respectively. Luminosities higher than $A \sim 0.7$ are not included, for avoiding the radiation pressure dominating over the attracting forces, and pushing the test particle to infinity.

Concerning the emitting surface, we fix its radius\footnote{The subsequent discussions are not affected if we choose $\RS\in(r_{\rm H},r_{\rm ps})$ \cite{Bakala2019}.} to $\RS=2.5M$ for both metrics and let its angular velocity vary as $\Omega_\star= (0,0.09,0.16)\ M^{-1}$. The maximum angular velocities are $\Omega_{\rm max}(2.5M)=0.179M^{-1}$ and $\Omega_{\rm max}(2.5M)=0.178M^{-1}$, corresponding to the Schwarzschild and to our BH solution, respectively. Therefore, we investigate the cases of static, medium, and high rotational spins. The parameters $A,\RS$, and $\Omega_\star$ can be inferred from X-ray observational data (see Refs.~\cite{Falanga2015,Fabian2015}, and references therein for details).  As the photon impact parameter \eqref{eq:explicit_impact_parameter} depends on the metric, we provide its expression both in the Schwarzschild geometry ($b_{\rm Sch}$) and in terms of our BH solution ($b_{\rm sol}$).
In our setup, if the critical region is located within the emission zone, then the test particle trajectory ends its motion when meeting the outermost external surface. The discrepancies among test particle trajectories are strongly enhanced by the relativistic PR effect due to its sensitive dependence on the initial data and the underlying geometry. Therefore, such a behaviour permits to single out particular features that can be surely detected through the data provided either by EHT or GRAVITY collaborations.

\subsection{Discussion on the  results}
\label{sec:discussion_PR}
We provided two examples of trajectories with different initial conditions on the test particle motion, namely: (1) same radius, but different velocity (see Fig.~\ref{fig:FigPRa}); (2) different radius and velocity (see Fig.~\ref{fig:FigPR}). We can immediately see that both simulations exhibit akin trajectories, although, in the second set,  differences appear more pronounced. For low luminosities, we note more similarities and this makes it difficult to appreciate possible deviations. In the case of low-intermediate luminosities, discrepancies are more visible for medium angular velocity. Finally, for high-intermediate luminosities, relevant differences emerge. 

It is important to note that  motion  of test particles can be mainly inferred by resorting to the ray-tracing technique. Depending on the matter type involved in the relativistic PR effect, we have: $(i)$ accreting plasma, captured by EHT data \cite{EHC20191,Akiyama2022image}; $(ii)$ astrophysical objects similar to  stars orbiting around Sgr A$^\star$ tracked by Gemini, Subaru, and VLT data provided by the GRAVITY collaboration (see \emph{e.g.}, Ref.~\cite{Do2019}, and references therein, for details).

From our simulations, intermediate luminosities, medium and high rotational angular velocities show that the considered BH geometry is different from the Schwarzschild one. For instance, setting $A=0.6$ and $\Omega_\star=0.16$, the test particle reaches the critical surface in the case of Schwarzschild metric, while it escapes to infinity for our BH solution. 

\section{Conclusions and perspectives}
\label{sec:end}
In this paper, we considered $f(R)$ theories of gravity in a static and spherically symmetric spacetime. By means of  Noether symmetries, we selected viable power-law models of the form $f(R) \sim R^k$. We found that the metric solutions to the field equations can be parameterized by two real constants ($\epsilon,\, r_0$), where $r_0$ is a constant and $\epsilon$ accounts for deviations from the Schwarzschild spacetime, being $k\equiv 1+\epsilon$, with $|\epsilon| \ll 1$. Physical considerations constrain the parameter $\epsilon$ in the range $- 1/2 \le \epsilon \le 0$. In particular, to detect small deviations from the Schwarzschild BH, we chose $\epsilon=-0.01$, as a  value that could lie under the current observational sensitivity.

We proposed a combination of different and known astrophysical techniques capable of detecting possible small metric departures from the Schwarzschild spacetime.
In particular, we considered two astrophysical scenarios: (1) gravity is the only force acting on the surrounding matter; (2) gravity and electromagnetic radiation processes alter the matter's motion.
The first case is analysed by exploiting: $(i)$ the BH image reconstruction via the ray-tracing technique; $(ii)$ the epicyclic frequencies; $(iii)$ the BH shadow profile. Our results demonstrate that the BH image can provide a deep and complete analysis of the underlying geometry. This is due to the extension of the image from the ISCO radius to large distances from the gravitational source.

A more complete and simple approach is represented by epicyclic frequencies. In this regard, the radial frequency is capable of highlighting differences with respect to the Schwarzschild metric, from the ISCO radius up to $r\sim7M$ and far from the gravitational source. Instead, the BH shadow profile cannot provide significant information, since it depends only on the critical impact parameter that, in our case, is extremely close to the Schwarzschild geometry. However, among all methods considered here, the last approach is the only one inquiring about gravity in strong regimes.

To summarize, these three astrophysical techniques result to be complementary and efficient in investigating BH physics and gravity. Indeed, they embrace a wide class of gravitational scenarios, ranging from strong (as the photon ring radius) to weak (as the outer edge of an accretion disk) regimes. In addition, such techniques also involve $g_{tt}$ and $g_{rr}$, thus being very sensitive to small departures from GR in both metric components. The current observational data can partially detect these metric departures, whereas the EHT data have to be upgraded in wavelength sensitivity.

Furthermore, we took into account the case where electromagnetic radiation processes are modelled via the relativistic PR effect. Since this phenomenon is very sensitive to the initial conditions \cite{Defalco2021chaos}, we first considered test particles with equal initial radii but different initial velocities, and then, with different initial radii and velocities. Even in this case, it is possible to inquire about gravity in various regimes and configurations. In particular, we found that the physical systems exhibiting more pronounced deviations from GR are those with intermediate luminosities and an emitting surface located close to the BH event horizon and rotating with either medium or high angular velocities. Exploring test particle trajectories in terms of the underlying relativistic PR parameters, it is possible to find particular configurations  identifying  the presence of metric changes, independently of the sensitivity of current observational EHT or GRAVITY data.

Thus, our study sheds light on the possibility to investigate metric departures via the synergy among various astrophysical techniques. Such methods could be applied to already existing data or very sensitive data from the near future. The primary goal of this work was to underline the predictive power of the above astrophysical techniques, which must be contextualized into the gravitational system under study for the sensibility and availability of the related observations. Certainly, additional and alternative strategies may be used to further constrain (and thus considerably reduce the degeneracy on) deviations from the Schwarzschild metric, such as gravitational waves, quasi-normal modes, and so forth. The methodology described here could be extended to axially symmetric and stationary spacetimes around uniformly spinning BHs. In the latter situation, the BH shadow profile can be useful to provide a more rich phenomenology, thus encoding valuable information on the gravitational background in the strong field regimes.

\section*{Acknowledgements} 
This paper is based upon work from COST Action CA21136 {\it Addressing observational tensions in cosmology with systematics and fundamental physics} (CosmoVerse) supported by European Cooperation in Science and Technology.
The authors acknowledge the Istituto Nazionale di Fisica Nucleare (INFN), Sezione di Napoli, \textit{iniziative specifiche} GINGER, TEONGRAV, QGSKY, and MOONLIGHT2. V.D.F. and S.C. acknowledge also the support of the Gruppo Nazionale di Fisica Matematica (GNFM) of Istituto Nazionale di Alta Matematica (INDAM). 

\section*{Data Availability Statement}
The data that support the findings of this study are available from the corresponding author upon reasonable request.

\end{document}